\pdfoutput=1
\documentclass[prx,twocolumn,showpacs,amsmath,amssymb,floatfix]{revtex4-1}

\usepackage{graphicx}  
\usepackage{dcolumn}   
\usepackage{bm}        
\usepackage{amssymb}   
\usepackage{amsmath}
\usepackage{mathrsfs}
\usepackage{epigraph}
\usepackage{braket}
\usepackage{gensymb}
\usepackage{lmodern}
\usepackage{lipsum}
\usepackage{marvosym}
\usepackage{ tipa }
\usepackage{bbold}
\usepackage{dsfont}
\usepackage{esint}
\usepackage{mathdots}
\usepackage{xcolor}
\usepackage{appendix}
\usepackage{natbib}
\usepackage{multirow}
\usepackage{float}
\usepackage{comment}
\usepackage[colorlinks=true, linkcolor=blue, urlcolor=blue,
citecolor=blue]{hyperref}

\DeclareMathOperator{\id}{\bold{\mathds{1}}}

\begin{document}
\title{Quantum many-body states and Green functions of nonequilibrium electron-magnon systems: Localized spin operators vs. their mapping to Holstein-Primakoff bosons} 	
\author{Utkarsh Bajpai}
\affiliation{Department of Physics and Astronomy, University of Delaware, Newark, DE 19716, USA}	
\author{Abhin Suresh}
\affiliation{Department of Physics and Astronomy, University of Delaware, Newark, DE 19716, USA}
\author{Branislav K. Nikoli\'{c}}
\email{bnikolic@udel.edu}
\affiliation{Department of Physics and Astronomy, University of Delaware, Newark, DE 19716, USA}

\begin{abstract}
The operators of localized spins within a  magnetic material commute at different sites of its lattice and anticommute on the same site, so they are neither fermionic nor bosonic operators. Thus, to construct diagrammatic many-body perturbation theory, requiring the Wick theorem,  the spin operators are usually mapped to the bosonic ones with Holstein-Primakoff (HP) transformation being the most widely used in magnonics and spintronics literature. However, to make calculations tractable, the square root of operators in the HP transformation is expanded into a Taylor series truncated to some low order. This poses a question on the {\em range of validity of truncated HP transformation} when describing nonequilibrium dynamics of localized spins interacting with each other or with  conduction electron spins---a problem frequently encountered in numerous transport phenomena in magnonics and spintronics. Here we apply exact diagonalization techniques to Hamiltonian of fermions (i.e., electrons) interacting with HP bosons vs. Hamiltonian of fermions interacting with the original localized spin operators in order to compare their many-body states and one-particle equilibrium or nonequilibrium Green functions. We employ as a test bed  a one-dimensional quantum Heisenberg ferromagnetic spin-$S$ XXX chain of $N \le 7$ sites, where $S=1$ or $S=5/2$, and the ferromagnet can be made  metallic by allowing electrons to hop between the sites while interacting with localized spin via $sd$ exchange interaction. For two different versions of the Hamiltonian for this model, we compare: the structure of their ground states; time evolution of excited states; spectral functions  computed from the retarded Green function in equilibrium; and the double-time-dependent lesser nonequilibrium Green function.  Interestingly, magnonic spectral function can be substantially modified, by acquiring additional peaks due to quasibound states of electrons and magnons, once the interaction between these subsystems is turned on. The Hamiltonian of fermions interacting with HP bosons gives incorrect ground state and electronic spectral function, unless large number of terms are retained in truncated HP transformation. Furthermore, tracking nonequilibrium dynamics of localized spins over longer time intervals requires progressively larger number of terms in truncated HP transformation even if small magnon density is excited initially, but the required  number of terms is reduced when interaction with conduction electrons is turned on. Finally, we show that  recently proposed [M.~Vogl \emph{et} \emph{al}., Phys. Rev. Research \textbf{2}, 043243 (2020); J.~K\"{o}nig \emph{et al}., SciPost Phys. \textbf{10}, 007 (2021)] resummed HP transformation, where spin operators are expressed as polynomials in bosonic operators, resolves the trouble with truncated HP transformation, while allowing us to derive an exact quantum many-body (manifestly Hermitian) Hamiltonian consisting of {\em finite and fixed} number of boson-boson and electron-boson interacting terms. 
\end{abstract}
\maketitle
\section{Introduction}\label{sec:intro}

The concept of spin waves was introduced by Bloch~\cite{Bloch1930} as a disturbance in the local magnetic ordering of ferromagnetic materials. In the spin wave, the expectation value of localized spin operators precess around the easy axis with the phase of precession of adjacent expectation values varying harmonically in space over the  wavelength $\lambda$. The quanta of energy of spin waves behave as quasiparticles termed magnons each of which carries energy $\hbar \omega$ and spin $\hbar$. 

As regards terminology, we note that in spintronics and magnonics~\cite{Chumak2015} literature it is common to use ``spin wave'' for excitations described by the classical Landau-Lifshitz-Gilbert (LLG) equation~\cite{Wieser2015} within numerical micromagnetics~\cite{Kim2010} or atomistic spin dynamics~\cite{Evans2014}, while ``magnon'' is used for quantized version of the same excitation. In other subfields of condensed matter physics, terms ``spin waves'' and ``magnons'' are sometimes used to distinguish between long- and short-wavelength excitations, respectively, or both names are used interchangeably~\cite{Zhitomirsky2013}.

\begin{figure}
	\centering
	\includegraphics[width=\linewidth]{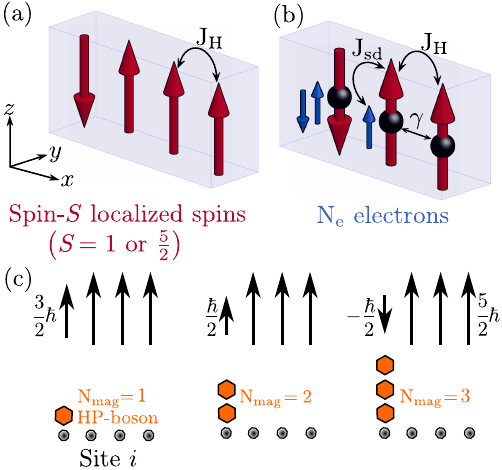}
	\caption{(a) Schematic view of a finite-size 1D quantum many-body system comprised of a chain of $N=4$ sites hosting spin-$S$ localized spins ($S=1$ or $S=\frac{5}{2}$ considered in this study) which interact with each other via the nearest-neighbor Heisenberg exchange interaction of strength $J_H$. At $t=0$, nonequilibrium dynamics can be initiated by flipping the localized spin on site $i=1$. The 1D quantum many-body system in (b) is the same as (a) but with smaller number of $N=3$ sites whose \mbox{spin-$S$} localized spins additionally interact with conduction electron spins (blue arrows) via the $sd$ exchange interaction of strength $J_\mathrm{sd}$. The conduction electrons hop between the sites with the hopping parameter \mbox{$\gamma$} where ``half-filled" (i.e., $N_e = 3$) tight-binding chain is used. Panel (c) illustrates the reduction of eigenvalues  
	$S^z_i$ of the $z$-component $\hat{S}^z_i$ of localized spin-$\frac{5}{2}$ operator on site $i=1$ by $N_\mathrm{mag}$ units, which is {\em equivalent} to creating $N_\mathrm{mag}$ Holstein-Primakoff bosons on site $i=1$ once the localized spins are mapped~\cite{Holstein1940} to bosonic operators.}
	\label{fig:fig1}
\end{figure}

The second-quantization description of magnons was introduced by Holstein and Primakoff (HP)~\cite{Holstein1940} by mapping the localized spin operator $\hat{\mathbf{S}}_i$ on site $i$ of the lattice to bosonic operators
\begin{subequations}\label{eq:hp}
	\begin{eqnarray}
	\hat{S}^+_i &=& \hat{S}^x_i + i\hat{S}^y_i  
	=  \sqrt{2S}\bigg(1 - \frac{\hat{n}_i}{2S}\bigg)^{1/2}\hat{a}_i, \label{eq:hp_sp}
	\\
	\hat{S}^-_i &=& \hat{S}^x_i - i\hat{S}^y_i 
	=  \sqrt{2S} \hat{a}_i^\dagger \bigg(1 - \frac{\hat{n}_i}{2S}\bigg)^{1/2}, \label{eq:hp_sm}
	\\
	\hat{S}^z_i &=&  (S - \hat{n}_i). \label{eq:hp_sz}
	\end{eqnarray}
\end{subequations}
Here $\hat{a}_i^\dagger$ ($\hat{a}_i$) creates (annihilates) HP boson on site $i$ and satisfies the bosonic commutation relations
\begin{equation}\label{eq:hp_comm}
[\hat{a}_i,\hat{a}_j^\dagger] = \id \delta_{ij},
\end{equation}
where $\id$ is the unit operator in the infinite dimensional Hilbert space of bosons. The HP boson number operator,  \mbox{$\hat{n}_i =\hat{a}_i^\dagger \hat{a}_i$} whose eigenvalues and eigenstates are defined by $\hat{n}_i |n\rangle = n |n\rangle$, measures how much the localized spin deviates away from the ground state [where the ferromagnetic ground state with the $z$-axis as the easy axis is assumed in Eq.~\eqref{eq:hp}]. Thus, the creation of one HP boson is equivalent to {\em removing of one unit} of spin angular momentum from the ground state [see Fig.~\ref{fig:fig1}(c) for illustration and Sec.~\ref{sec:magnons} for technical details].

The  textbook literature~\cite{Mahan2011,Chudnovsky2006} is typically focused on band structure of noninteracting magnons (which can also be topologically nontrivial~\cite{Kim2016,Mook2021}), so it discusses only the lowest-order truncation 
\begin{subequations}\label{eq:approxhp}
\begin{eqnarray}
\hat{S}^+_i  & \approx & \sqrt{2S}  \hat{a}_i, \\
\hat{S}^-_i & \approx  & \sqrt{2S} \hat{a}_i^\dagger, 
\end{eqnarray}
\end{subequations}
of the original HP transformation in Eq.~\eqref{eq:hp} while retaining the terms in the Hamiltonian that are up to the quadratic order in the bosonic operators. This effectively assumes low-density limit \mbox{$\langle \hat{n}_i \rangle/2S \ll 1$} achieved at, e.g., sufficiently low temperatures~\cite{Elyasi2020}  and/or large $S \gg 1$ in which HP bosons can be treated as noninteracting.  Taking into account higher order terms in the Hamiltonian generated by Eq.~\eqref{eq:approxhp}, as well as in the Taylor expansion of the square root in Eq.~\eqref{eq:hp}, produces higher-than-quadratic terms in the bosonic operators which describe {\em boson-boson interactions}~\cite{Zhitomirsky2013,Chudnovsky2006,Mook2021,Tupitsyn2008,Yuan2020,Takei2019,Elyasi2020} leading to renormalization of magnon energy, magnon decay (one magnon decays into two)~\cite{Zhitomirsky2013}, coalescence (two magnons coalesce into one), four-magnon interactions, decays into four magnons and other higher order processes~\cite{Radosevic2015}. 

Since bosonic operators $\hat{a}_i^\dagger$, $\hat{a}_i$  act on an {\em infinite}-dimensional Hilbert space, but the physical Hilbert space corresponding to
a single localized spin on site $i$ is spanned by only $2S + 1$ states, the extra unphysical states are decoupled from the physical ones by the square root in Eq.~\eqref{eq:hp}. Such exact HP transformation in Eq.~\eqref{eq:hp} splits the infinite dimensional Hilbert space spanned by boson number states $\{ |n\rangle \}_{n \in \mathbb{N}}$ into two sectors---physical states $\{ |n\rangle \}_{n=0,...,2S}$; and all the unphysical ones $\{ |n\rangle \}_{n>2S}$ [see also Eq.~\eqref{eq:phys_states}]. Those sectors {\em cannot} be connected by $\hat{S}^+_i$ and $\hat{S}^-_i$ operators. However, when the square root in Eq.~\eqref{eq:hp} is expanded in power series and then truncated (see Secs.~\ref{sec:hptruncated} and~\ref{sec:hpresum}) to any {\em finite} order $N_T$, the physical and unphysical subspaces become coupled. In addition, canonical commutation relations for the spin operators are then satisfied only approximately, resulting in artificial breaking of rotational symmetries that may be present in the original Hamiltonian~\cite{Vogl2020,Konig2021}.

Retaining higher order terms in the Taylor series expansion of Eq.~\eqref{eq:hp} is necessary to study, e.g., equilibrium properties at increasing temperature~\cite{Dyson1956,Hofmann2011,Radosevic2015} or  nonequilibrium dynamics~\cite{Tupitsyn2008,Yuan2020,Schuckert2018,Takei2019}.  For example, Dyson~\cite{Dyson1956} calculated how magnetization of the Heisenberg model of a three-dimensional ferromagnet decays with temperature, $M(T)/M(0)=1 - c_0 T^{3/2} - c_1 T^{5/2} - c_2 T^{7/2} - c_3 T^4 + \mathcal{O}(T^{9/2})$, where $T^{3/2}$ is the so-called Bloch law for noninteracting magnons with parabolic energy-momentum dispersion; second and third term also stem from noninteracting magnons but with nonparabolic dispersion on a discrete lattice; and magnon-magnon interactions start manifesting at order $T^4$. Such calculations require many-body perturbation theory (MBPT)~\cite{Stefanucci2013,Schlunzen2020} which is most easily formulated in terms of bosonic or fermionic operators. For such  operators, the Wick theorem~\cite{Leeuwen2012} for their averages  over the noninteracting system makes it possible to expand properties of the interacting system into the Feynman  diagrammatic series of perturbation order $\mathcal{O}(g^n)$ where $g$ is the strength of fermion-boson interaction. On the other hand, spin operators which commute on different sites and anticommute on the same site are neither fermionic nor bosonic operators, and there is no  Wick theorem for spin operators.

Since higher order terms in the power-series expansion of square root in the HP transformation in Eq.~\eqref{eq:hp}, conjectured by Kubo~\cite{Kubo1953} to be  only an asymptotic series, lead to cumbersome MBPT~\cite{Harris1971,Hamer1992}, a plethora of other mappings of original localized spin operators to bosonic or fermionic operators has been proposed. For example, one has a choice to map to Dyson-Maleev bosons~\cite{Dyson1956,Maleev1958}, Schwinger bosons~\cite{Schuckert2018}, fermions~\cite{Jordan1928,Affleck1998}, Majorana fermions~\cite{Tsvelik1992}, sypersymmetric operators~\cite{Coleman2000} and  exotic particles called semions~\cite{Kiselev2000}. We note that mapping of localized spin operators to bosonic or fermionic operators can be evaded altogether  within the path integral formulation by using spin coherent states, but that approach leads to topological terms associated with the Berry phase so that even in path integral formalism mapping to bosonic operators is preferred~\cite{Schuckert2018}. Although the Dyson-Maleev transformation evades usage of the square root of operators in Eq.~\eqref{eq:hp}, it generates Hamiltonian that is {\em no longer manifestly Hermitian}. The Schwinger transformation require the introduction of auxiliary fields. The Jordan-Wigner transformation~\cite{Jordan1928} or mapping to Majorana fermions~\cite{Tsvelik1992} are exact, but they work only for $S=1/2$ operators.  

These drawbacks  have prompted very recent reexaminations~\cite{Vogl2020,Konig2021} of HP transformation to find possible nonperturbative replacements of the Taylor series of the square root in Eq.~\eqref{eq:hp} which can be written as a polynomial in bosonic operators, while ensuring no coupling between physical and unphysical subspaces as well as  {\em manifestly Hermitian} bosonic Hamiltonian.  Although such polynomial expressions do not reproduce exactly the canonical commutation relations for the spin operators, the extra terms generated turn out to be unimportant because they do not couple physical and unphysical subspaces of the bosonic Hilbert space, i.e., they act solely on the unphysical subspace~\cite{Vogl2020, Konig2021}. 

The MBPT calculations of equilibrium magnon properties based on Dyson-Maleev vs. truncated HP transformation have been carefully compared in the literature over many decades~\cite{Harris1971,Hamer1992}. On the other hand, much less is know about the range of validity~\cite{Marcuzzi2016,Hirsch2013} of truncated HP transformation when describing nonequilibrium dynamics of localized spins, including situations where {\em additional interactions with conduction electrons} are  present. The electron--localized-spin interactions are frequently encountered in quantum transport phenomena in spintronics. The nonequilibrium MBPT~\cite{Stefanucci2013,Schlunzen2020} for such problems is virtually always conducted using truncated HP transformation, as exemplified by theoretical and computational modeling of inelastic electron tunneling spectroscopy in magnetic tunnel junctions~\cite{Mahfouzi2014}; spin-transfer~\cite{Tay2013,Cheng2019,Bender2019} and spin-orbit torques~\cite{Okuma2017}; ultrafast demagnetization~\cite{Tveten2015}; and conversion of  magnonic spin currents into electronic spin current (or vice versa) at magnetic-insulator/normal-metal interfaces~\cite{Zheng2017,Troncoso2019,Adachi2011}. Similarly, truncated HP transformation is typically chosen for problems in quantum magnonics, such as for  nonequilibrium dynamics of localized spins within magnetic insulators~\cite{Tupitsyn2008,Yuan2020,Takei2019,Kamra2016}; their interaction with external electromagnetic fields~\cite{Parvini2020,Elyasi2020}; 
and analysis of coherence of magnon quantum states~\cite{Yuan2020,Bender2019a}. One can expect that truncated HP transformation will eventually break down at sufficiently long times (as confirmed Figs.~\ref{fig:fig4} and ~\ref{fig:fig5}) when higher-order terms in the expansion of the square root in HP transformation become important. Such breakdown then precludes~\cite{Tay2013} accurate tracking of nonequilibrium dynamics of localized spins, which can be driven far from their initial direction (along the easy axis) and eventually reversed by, e.g., spin-transfer torque~\cite{Ralph2008,Petrovic2021}. Surprisingly, {\em rigorous analysis of such breakdown is lacking}. 

Instead, current-driven magnetization reversal via spin-transfer torque~\cite{Ralph2008} is standardly modeled by the LLG equation~\cite{Berkov2008}, which is  combined in a  multiscale  fashion with some type of steady-state~\cite{Ellis2017} or time-dependent quantum transport calculations~\cite{Petrovic2018,Bajpai2019a,Suresh2020,Suresh2021,Bajpai2020,Stahl2017,Bostrom2019} considering single-particle quantum Hamiltonians for electrons. Thus, such hybrid quantum-classical theories are justified only in the classical limit $\hbar \rightarrow 0$ and for large localized spins $S \rightarrow \infty$ (while $S \times \hbar  \rightarrow 1$)~\cite{Wieser2015,Stahl2017,Gauyacq2014}, as well as in the absence of entanglement~\cite{Wieser2015,Mondal2019,Petrovic2021} between quantum states of localized spins. For example, in the emerging concept of quantum spin torque~\cite{Mondal2019,Petrovic2021,Petrovic2021a,Mitrofanov2020,Mitrofanov2021}, describing transfer of angular momentum between spins of flowing electrons and localized spins in situations~\cite{Zholud2017} where the latter {\em must} be described by quantum-mechanical operators, the whole system of electrons and localized spins can only be modeled by a quantum many-body Hamiltonian [as exemplified by Eqs.~\eqref{eq:ham_sys} and ~\eqref{eq:em_ham}]. 

In this study, we apply exact diagonalization techniques~\cite{Wang2019} to quantum many-body Hamiltonians defined on a one-dimensional (1D) chain of  $N$ sites hosting fermionic (for electrons) and localized spin operators, or fermionic and bosonic (obtained by mapping the original localized spin operators) operators. By comparing their many-body quantum states and Green functions (GFs), both in {\em equilibrium} and in {nonequilibrium}, makes it possible to precisely delineate {\em the range of validity of truncated HP transformation}. We consider 1D quantum Heisenberg ferromagnetic spin-$S$ XXX chain hosting localized spins which interact via the nearest-neighbor exchange interaction of strength $J_H$, as illustrated in Fig.~\ref{fig:fig1}(a), where both spin $S=1$ (as the ``ultraquantum'' limit) and $S=5/2$ (as in, e.g., Fe$^{3+}$ valence state with five 3d electrons coupled by Hund's rule into the high spin state forming a localized $S = 5/2$ moment) are employed. Naively, the eigenvalue of $\hat{\mathbf{S}}_i^2$ being $S^2(1+1/S)$ suggests that quantum effects become progressively less important for $S>1$, but they exist for all $S < \infty$ vanishing as $1/2S$ in the classical limit~\cite{Parkinson1985}. The nonzero electron hopping $\gamma$ between the sites, where $N=3$ sites are chosen when electrons are present as illustrated in Fig.~\ref{fig:fig1}(b), means that such 1D chain models a ferromagnetic metal (FM). Its $N_e$  conduction electrons [we consider half filled lattice, so $N_e=3$ for systems in Fig.~\ref{fig:fig1}(b)] interact with localized spins via $sd$ exchange interaction~\cite{Cooper1967} usually considered in spintronics. From the viewpoint of the physics of strongly correlated electrons, the model illustrated in Fig.~\ref{fig:fig1}(b) can also be interpreted as the Kondo-Heisenberg chain~\cite{Tsvelik2017}. 

For technical reasons (i.e., exponential increase of the size of matrix representation of Hamiltonian), we consider 1D chains of $N \le 7$ sites while concentrating on {\em generic features} which are not bound to one dimension or small number of electrons and localized spins considered. In fact, artificial atomic chains have also been realized experimentally using  ferromagnetically~\cite{Spinelli2014} or antiferromagnetically~\cite{Loth2012} coupled few Fe atoms on a substrate, where magnons along the chain were excited and detected via atom-resolved inelastic tunneling spectroscopy in a scanning tunneling microscope~\cite{Spinelli2014}. 

It is also worth recalling that small clusters (composed of, e.g., 2--8 lattice sites) in 1, 2, and 3 spatial dimensions---hosting electrons interacting with each other via the on-site or nearest-neighbor Coulomb interaction~\cite{Schumann2010,Carrascal2015,Hermanns2014} (as described by  ``pure'' and extended Hubbard models~\cite{Schumann2010}, respectively); or electrons interacting with bosons~\cite{Sakkinen2015,Sakkinen2015a,Dimitrov2017}---have played an important role in testing approximation schemes for quantum many-body problem against numerically exact benchmarks in different subfields of condensed matter and atomic-molecular-optical physics. Furthermore, the advent of numerically exact algorithms and supercomputers has led to recent re-examination of many physically motivated simplifications and approximations developed earlier in quantum many-body theory for condensed matter systems (such as Migdal-Eliashberg theory for electron-phonon systems~\cite{Esterlis2018}; partial summation of classes in Feynman diagrams in MBPT~\cite{Gukelberger2015}; and existence of Luttinger-Ward functional of dressed one-particle Green function~\cite{Kozik2015}) in order to draw boundaries of parameters for which their complete breakdown ensues. Our study proceeds in the same spirit, where we {\em explicitly delineate ``breakdown'' times}---in Fig.~\ref{fig:fig4} for pure localized spins  and in Fig.~\ref{fig:fig5} for localized spins interacting with conduction electrons---at which widely used in spintronics and magnonics truncated versions of the HP transformation in Eq.~\eqref{eq:hp} inevitably break down by generating  quantum time evolution which starts to substantially deviate from the exact one obtained by using the original localized spin operators.  

The paper is organized as follows. In Sec.~\ref{sec:theory} we introduce different versions of quantum many-body Hamiltonian describing systems in Fig.~\ref{fig:fig1}  and their matrix representations, as well as a procedure to obtain the exact one-particle double-time-dependent retarded and lesser GFs. In particular, subsection~\ref{sec:hptruncated}  introduces an infinite power series expansion of the HP transformation from Eq.~\eqref{eq:hp} and its truncation, while subsection~\ref{sec:hpresum} provides a brief summary of recently proposed~\cite{Vogl2020, Konig2021} resummation of  truncated HP transformation. The time evolution of quantum many-body states of a spin chain with no electrons ($N_e=0$) is employed in Sec.~\ref{sec:validity_mm} to examine the range of validity of truncated HP transformation when tracking time evolution of localized spins in the presence of \emph{magnon-magnon} interaction and different number of initially excited magnons $N_\mathrm{mag}$. Then in Sec.~\ref{sec:validity_em} we introduce electrons into 1D chain to examine the range of validity of truncated HP transformation when tracking time evolution of localized spins in the presence of {\em both}  magnon-magnon and electron-magnon interactions. In the same Sec.~\ref{sec:validity_em}, we additionally employ resummation~\cite{Vogl2020, Konig2021} of truncated HP transformation to derive quantum many-body Hamiltonian [Eq.~\eqref{eq:em_ham}] for electron-magnon systems in terms of fermionic and bosonic operators whose usage reproduces numerically exact result from calculations based on the original localized spin operators. In Secs.~\ref{sec:groundstate} and ~\ref{sec:spectralelectron} we compare ground state and electronic spectral function (or ``interacting density of states''~\cite{Balzer2011,Nocera2018}) of quantum many-body Hamiltonian in terms of the original localized spin operators vs. Hamiltonian using bosonic operators generated by truncated HP transformation. The  magnonic spectral function and related excited eigenstates are  studied in Sec.~\ref{sec:spectralmagnon}. Since both ground and excited states of electron-magnon interacting system are many-body entangled~\cite{Chiara2018}, we compute their entanglement entropy in Sec.~\ref{sec:entanglement} which makes it possible to quantify how far they are from the eigenstates of a system where the interaction between electrons and localized spins is turned off. Finally, Sec.~\ref{sec:diag_offdiag} studies time evolution of diagonal and off-diagonal elements of time-dependent lesser electronic and magnonic GFs which demonstrates that often employed ``local self-energy approximation''~\cite{Luiser2009,Rhyner2014,Cavassilas2016,Bescond2018} for electron-boson interacting systems, neglecting the off-diagonal elements, is generally not justified. We conclude in Sec.~\ref{sec:conclusions}.

 \section{Models and Methods}\label{sec:theory}
\subsection{Quantum many-body Hamiltonian of electrons interacting with localized spins}\label{sec:hamilt}
The quantum many-body Hamiltonian of 1D chain composed of $N$ sites (with open boundary conditions assumed), each of which hosts spin-$S$ localized spin which interacts with conduction electron spins [as illustrated in Fig.~\ref{fig:fig1}(b)], is given by~\cite{Woolsey1970}
\begin{equation}\label{eq:ham_sys}
\hat{H} = \hat{H}_e + \hat{H}_\mathrm{lspins} + \hat{H}_\mathrm{e-lspins}.
\end{equation}
It acts in the total space $\mathcal{F}_e \otimes \mathcal{H}_\mathrm{lspins}$ which is a tensor product of the Fock space of electrons, $\mathcal{F}_e$, and the Hilbert space of all localized spins 
\begin{equation}\label{eq:hlspins}
\mathcal{H}_\mathrm{lspins} = \mathcal{H}_1 \otimes \cdots \otimes \mathcal{H}_N.
\end{equation}
The Fock space of electrons~\cite{Schlunzen2020}  
\begin{equation}\label{eq:fock}
   \mathcal{F}_e= \overline{\mathbb{C} \oplus  \mathcal{H}_e \oplus \hat{\mathcal{A}}(\mathcal{H}_e \otimes \mathcal{H}_e)  \oplus  \hat{\mathcal{A}}(\mathcal{H}_e \otimes \mathcal{H}_e  \otimes \mathcal{H}_e) \oplus \cdots}, 
\end{equation}
is induced by the one-electron Hilbert space $\mathcal{H}_e$ as the completion (indicated by overline) of the direct sum of
antisymmetrized $n$-fold tensor products of $\mathcal{H}_e$. The operator $\hat{\mathcal{A}}$ antisymmetrizes tensors for fermionic particles. In the sector of $\mathcal{F}_e \otimes \mathcal{H}_\mathrm{lspins}$ with $N_e=0$ electrons, we have a chain hosting only spin-$S$ localized spins [as illustrated in Fig.~\ref{fig:fig1}(a)], which is described solely by $\hat{H}_\mathrm{lspins}$ term 
\begin{equation}\label{eq:spin_ham}
\hat{H}_\mathrm{lspins} = -J_H \sum_{\braket{ij}} \hat{\bold{S}}_i\cdot \hat{\bold{S}}_j,  
\end{equation}
chosen as the quantum Heisenberg Hamiltonian with the nearest-neighbor (NN) exchange interaction (as signified by $\braket{ij}$ notation) of strength \mbox{$J_H=1$ eV}. When electrons are present, they are described by $\hat{H}_e$ term 
\begin{equation} \label{eq:elec_ham}
	\hat{H}_e = -\gamma \sum_{\braket{ij}} \hat{\psi}_i^\dagger \hat{\psi}_j,
\end{equation}
chosen as the tight-binding Hamiltonian with the NN hopping \mbox{$\gamma=1$ eV} between single $s$-orbitals residing on each site. The Hamiltonian describing $sd$ exchange interaction of strength \mbox{$J_\mathrm{sd}=0.2$ eV}~\cite{Cooper1967} between conduction electron spin and localized spins is given by 
\begin{equation}\label{eq:sd_ham}
\hat{H}_\mathrm{e-lspins} = -J_\mathrm{sd} \sum_{i=1}^{N} \hat{\psi}_i^\dagger \hat{\bm \sigma} \hat{\psi}_i  \cdot \hat{\bold{S}}_i. 
\end{equation}
The row vector operator $\hat{\psi}_i^\dagger = (\psi_{i\uparrow}^\dagger,\psi_{i\downarrow}^\dagger )$ consists of operators $\hat{\psi}_{i\sigma}^\dagger$ which create an electron of spin $\sigma = \uparrow, \downarrow$ on site $i$; $\hat{\psi}_i$ is a column vector operator that contains the corresponding annihilation operators; and  $\hat{\bm \sigma} = (\hat{\sigma}^x,\hat{\sigma}^y,\hat{\sigma}^z)$ is the vector of the $2\times2$ Pauli spin matrices as matrix representation of spin-$\frac{1}{2}$ operator of electronic spin.

Using notation $\{\hat{O}_1,\hat{O}_2\}$ for the anticommutator and $[\hat{O}_1,\hat{O}_2]$ for the commutator of two operators $\hat{O}_1$ and $\hat{O}_2$, fermionic operators of electrons satisfy
\begin{equation}\label{eq:fermi_comm}
\{ \hat{\psi}_{i\sigma},\hat{\psi}^\dagger_{j\sigma'}\} = \id \delta_{ij}\delta_{\sigma\sigma'},
\end{equation}
where $\id$ is $4^N \times 4^N$ unit matrix in the antisymmetrized $2N$-particle subspace of the Fock space  $\mathcal{F}_e$. The localized spin operators $\hat{S}^\alpha_i$ ($\alpha = x,y,z$) on site $i$ satisfy the angular momentum algebra 
\begin{subequations}\label{eq:ang_al}
\begin{eqnarray}
&[\hat{S}^x_a,\hat{S}^y_b] = i \hat{S}_b^z\delta_{ab}, \label{eq:ang_al_a}
\\
&[\hat{S}^y_a,\hat{S}^z_b] = i \hat{S}_b^x\delta_{ab}, \label{eq:ang_al_b}
\\
&[\hat{S}^z_a,\hat{S}^x_b] = i \hat{S}_b^y\delta_{ab}. \label{eq:ang_al_c}
\end{eqnarray}
\end{subequations}
The square of the localized spin operator,  $\hat{S}^2_i = (\hat{S}^x_i)^2 + (\hat{S}^y_i)^2 + (\hat{S}^z_i)^2$, commutes with each component 
\begin{equation}\label{eq:tot_ang}
[\hat{S}^2_i,\hat{S}^\alpha_j] = 0.
\end{equation}
For computational convenience in calculations of electronic GFs, we change~\cite{Frederiksen2004} the basis of one-particle electronic states from site basis to eigenenergy basis to obtain
\begin{equation}\label{eq:eham_diag}
\hat{H}_e = \sum_{i=1}^{N} \epsilon_{i}\hat{c}^\dagger_{i} \hat{c}_{i},
\end{equation} 
where $c_i^\dagger = (c_{i\uparrow}^\dagger,c_{i\downarrow}^\dagger )$ is a row vector consisting of $c^\dagger_{i\sigma}$ operators which create an electron  with spin $\sigma = \uparrow, \downarrow$ in one-particle electronic eigenstate $\ket{\epsilon_i}$ with the discrete eigenenergy $\epsilon_{i}$, so that \mbox{$\hat{H}_e \ket{\epsilon_{i}} = \epsilon_{i}\ket{\epsilon_{i}}$}. These eigenenergies and eigenstates are evaluated by diagonalizing the one-particle tight-binding Hamiltonian 
\begin{equation}\label{eq:ham_1ele}
\hat{H}_e = \sum_{\braket{ij}}-\gamma \ket{i}\bra{j}, 
\end{equation}
where $\ket{i}$ denotes $s$-orbital [whose coordinate representation is $\langle \mathbf{r}| i \rangle = \phi(\mathbf{r}-\mathbf{R}_i)$] of an electron centered  on site $i$. Using change of basis transformation rules for operators in second-quantization formalism
\begin{equation}\label{eq:cc_dag_trans}
\hat{\psi}_{i\sigma} = \sum_{j=1}^N \braket{i|\epsilon_j}\hat{c}_{j\sigma}, 
\end{equation}
and substituting this into Eq.~\eqref{eq:sd_ham} we get
\begin{equation}\label{eq:sd_new_bas}
\hat{H}_\mathrm{e-lspins} = -J_\mathrm{sd} 
\sum_{i=1}^N\sum_{j=1}^N\sum_{j'=1}^N
\braket{\epsilon_{j'}|i}\braket{i|\epsilon_j}
\hat{c}_{j'}^\dagger \hat{\bm \sigma} \hat{c}_j \cdot \hat{\mathbf{S}}_i.
\end{equation}
Since each $\hat{c}_i^\dagger$ or $\hat{c}_i$ operator is represented by $4 \times 4$ matrix (see Sec.~\ref{sec:matrixofccdagger}), and each $\hat{\mathbf{S}}_i$ operator is represented by a $(2S+1) \times (2S+1)$ matrix, the quantum many-body Hamiltonian in Eq.~\eqref{eq:ham_sys} for the chain of $N$ sites in Fig.~\ref{fig:fig1}(b) is represented by a matrix of size $[4^N \times (2S+1)^N] \times [4^N \times (2S+1)^N]$. Although systems containing larger than our choice  $N=3$ (when electrons are present) or $N \le 7$ (when electrons are absent) sites could be diagonalized with state-of-the-art numerical algorithms~\cite{Wang2019}, we restrict our analysis to such smaller number of sites in order to make the analysis transparent and pedagogical---see, e.g., easy-to-follow visualization of ground and excited quantum many-body states depicting population of a small number of energy levels $\epsilon_i$ in Figs.~\ref{fig:fig7} and ~\ref{fig:fig9}, respectively.

\subsection{Symmetries of quantum many-body Hamiltonian}
%
\begin{figure*}
    \centering
    \includegraphics[width=\linewidth]{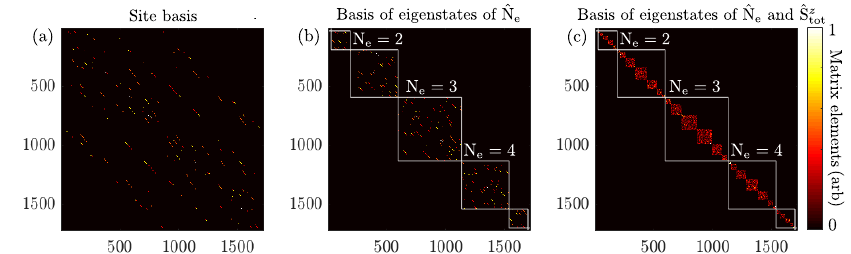}
    \caption{Visualization of the absolute value of matrix elements of Hamiltonian $\hat{H}$ [Eq.~\eqref{eq:ham_sys}] for 1D quantum many-body system of $N_e$  conduction electrons interacting with localized spins $S=1$ illustrated in Fig.~\ref{fig:fig1}(b). The matrix representation of $\hat{H}$ is given: (a) in site basis; (b) in basis composed of eigenstates of electron number operator $\hat{N}_e$ [Eq.~\eqref{eq:ele_num}], so that each block consists of states with fixed number of electrons $N_e$ ($N_e = 2, 3, 4$ is marked explicitly); and (c) in basis composed of eigenstates of \emph{both} $\hat{N}_e$  [Eq.~\eqref{eq:ele_num}] and total $z$-spin operator $\hat{S}^z_\mathrm{tot}$  [Eq.~\eqref{eq:total_sz}].}
    \label{fig:fig2}
\end{figure*}
The exact diagonalization of quantum many-body Hamiltonian in Eq.~\eqref{eq:ham_sys}
\begin{equation}\label{eq:ham_spec}
\hat{H} \ket{\Psi_k} = E_k \ket{\Psi_k},
\end{equation}
yields its many-body eigenenergies $E_k$ and many-body eigenstates $\ket{\Psi_k}$. The total electron number operator is given by
\begin{equation}\label{eq:ele_num}
\hat{N}_e = \sum_{i=1}^{N} (\hat{c}_{i\uparrow}^\dagger \hat{c}_{i\uparrow} + \hat{c}_{i\downarrow}^\dagger \hat{c}_{i\downarrow}).
\end{equation}
The operator of total spin in the $z$-direction
\begin{equation}\label{eq:total_sz}
\hat{S}^z_\mathrm{tot} = \frac{1}{2}\sum_{i=1}^N(\hat{c}_{i\uparrow}^\dagger \hat{c}_{i\uparrow} - \hat{c}_{i\downarrow}^\dagger \hat{c}_{i\downarrow}) + \sum_{i=1}^N \hat{S}^z_i,
\end{equation}
is the sum of electronic spin operators (first term) and localized spin operators (second term) along the $z$-axis at each site $i$. The many-body Hamiltonian in Eq.~\eqref{eq:ham_sys} has two symmetries encoded by the commutation relations:
\begin{equation}\label{eq:num_sym}
[\hat{H},\hat{N}_e] = 0, 
\end{equation}
which is due to conservation of the number of electrons $N_e$; and 
\begin{equation}\label{eq:sz_sym}
[\hat{H},\hat{S}_\mathrm{tot}^z] = 0, 
\end{equation}
which is due to conservation of total $z$-spin (electronic + localized spin) $S^z_\mathrm{tot}$. Therefore, $N_e$ and $S^z_\mathrm{tot}$, as eigenvalues of $\hat{N}_e$ and $\hat{S}_\mathrm{tot}^z$, respectively, serve as ``good quantum numbers" for labeling quantum many-body eigenstates
\begin{equation}\label{eq:qnumbers}
\ket{\Psi_k} = \ket{E_k, N_e, S^z_\mathrm{tot}}, 
\end{equation}
together with many-body eigenenergy $E_k$. 

The effect of two symmetries in Eq.~\eqref{eq:num_sym} and Eq.~\eqref{eq:sz_sym} can also be visualized in the matrix representation (see Secs.~\ref{sec:matrixofccdagger} and~\ref{sec:exact}) of quantum many-body Hamiltonian $\hat{H}$~[Eq.~\eqref{eq:ham_sys}]. For example, in Fig.~\ref{fig:fig2}(a)  the matrix elements of $\hat{H}$ in the original site basis are visually scattered throughout the whole matrix. However,  when $\hat{H}$ is represented in the basis of eigenstates of $\hat{N}_e$, Fig.~\ref{fig:fig2}(b) shows that its matrix becomes block-diagonal where each block contains the nonzero matrix elements associated with states with fixed number of electrons $N_e$. Finally, in Fig.~\ref{fig:fig2}(c) $\hat{H}$ is represented in the basis composed of eigenstates of $\hat{N}_e$ and  $\hat{S}^z_\mathrm{tot}$ simultaneously, which isolates additional submatrices with fixed $S^z_\mathrm{tot}$  within blocks associated to fixed $N_e$.

\subsection{Matrix representation of electronic creation and annihilation operators}\label{sec:matrixofccdagger}
A fermionic operator creating or annihilating electrons on a single site operate within the natural basis of kets  $\ket{0}$, $\ket{\uparrow}$, $\ket{\downarrow}$ and $\ket{\uparrow\downarrow}$ which denote the empty state; state with one \mbox{spin-$\uparrow$} electron; state with one \mbox{spin-$\downarrow$} electron; and the state with one \mbox{spin-$\uparrow$} and one \mbox{spin-$\downarrow$} electron. Thus, these basis states are represented by column vectors
\begin{equation}\label{eq:electron_basis}
\ket{0} = 
\begin{pmatrix}
1 \\
0 \\
0 \\
0 \\
\end{pmatrix} \hspace{0.1cm},\hspace{0.1cm}
\ket{\uparrow} = 
\begin{pmatrix}
0 \\
1 \\
0 \\
0 \\
\end{pmatrix} \hspace{0.1cm},\hspace{0.1cm}
\ket{\downarrow} = 
\begin{pmatrix}
0 \\
0 \\
1 \\
0 \\
\end{pmatrix} \hspace{0.1cm},\hspace{0.1cm}
\ket{\uparrow \downarrow} = 
\begin{pmatrix}
0 \\
0 \\
0 \\
1 \\
\end{pmatrix}.
\end{equation}
In the same basis, creation and annihilation operators that act in the 1-site or 2-particle subspace \mbox{$\mathscr{H}_{N=1} = \hat{\mathcal{A}}(\mathcal{H}_e \otimes \mathcal{H}_e)$} of the Fock space $\mathcal{F}_e$ are represented by $4 \times 4$ matrices
\begin{equation}
\hat{\psi}^\dagger_{\uparrow} = \begin{pmatrix}
0 & 0 & 0 & 0 \\
1 & 0 & 0 & 0 \\
0 & 0 & 0 & 0 \\
0 & 0 & 1 & 0
\end{pmatrix} \hspace{0.1cm},\hspace{0.1cm}
\hat{\psi}^\dagger_{\downarrow} = \begin{pmatrix}
0 & 0 & 0 & 0 \\
0 & 0 & 0 & 0 \\
1 & 0 & 0 & 0 \\
0 & -1 & 0 & 0
\end{pmatrix} \hspace{0.1cm},\hspace{0.1cm}
\hat{\psi}_{\sigma} = [\hat{\psi}^\dagger_{1\sigma}]^T,
\end{equation}
which satisfy the fermionic commutation relations in Eq.~\eqref{eq:fermi_comm}.
If we consider two sites, then electronic creation (annihilation) operators,  $\hat{\psi}_1^\dagger$ ($\hat{\psi}_1$) and $\hat{\psi}_2^\dagger$ ($\hat{\psi}_2$), act in the 2-site or 4-particle subspace \mbox{$\mathscr{H}_{N=2} = \hat{\mathcal{A}}(\mathcal{H}_e \otimes \mathcal{H}_e \otimes \mathcal{H}_e \otimes \mathcal{H}_e)$} of the Fock space $\mathcal{F}_e$  and are represented by matrices of size $4^2 \times 4^2$. For example, the action of $\hat{\psi}^\dagger_{1\sigma}$ in $\mathscr{H}_{N=2}$ is given by  
\begin{equation}
(\hat{\psi}^\dagger_{1\sigma})_{\mathscr{H}_{N=2}} = \hat{\psi}^\dagger_{\sigma}\otimes \id,  
\end{equation}
where $\id$ is  4 $\times$ 4 unit matrix. However, the action of $(\psi^\dagger_{2\sigma})_{\mathscr{H}_{N=2}}$ 
\begin{equation}\label{eq:pmatrix}
(\psi^\dagger_{2\sigma})_{\mathscr{H}_{N=2}} = \hat{P} \otimes \hat{\psi}^\dagger_\sigma.
\end{equation}
requires~\cite{Frederiksen2004} the permutation matrix \mbox{$\hat{P}=\text{diag}(1, -1, -1, 1)$}, instead of na\"ively using only the unit matrix $\id$, in order to preserve the correct anticommutation relations of fermionic operators at different sites in Eq.~\eqref{eq:fermi_comm}. The next step is to construct the matrix representation of  electronic creation and annihilation operators for three sites, which is done in a similar fashion~\cite{Frederiksen2004} to furnish 
\begin{subequations}\label{eq:ele_create}
\begin{eqnarray}
(\hat{\psi}_{1\sigma}^\dagger)_{\mathscr{H}_{N=3}} &=& \psi^\dagger_\sigma \otimes \id \otimes \id, \\ 
(\hat{\psi}_{2\sigma}^\dagger)_{\mathscr{H}_{N=3}} &=& \hat{P} \otimes \hat{\psi}^\dagger_\sigma \otimes \id, \\
(\hat{\psi}_{3\sigma}^\dagger)_{\mathscr{H}_{N=3}}  &=&  \hat{P} \otimes \hat{P} \otimes \hat{\psi}^\dagger_\sigma, 
\end{eqnarray}
\end{subequations}
where each operator on the left hand side (LHS) is a $4^3 \times 4^3$ matrix. Equations~\eqref{eq:ele_create} also make it clear how to construct inductively matrix representations of electronic creation and annihilation operators for arbitrary number of sites $N$, where these operators act in  \mbox{$\mathscr{H}_N = \hat{\mathcal{A}}(\underbrace{\mathcal{H}_e \otimes \mathcal{H}_e  \cdots \mathcal{H}_e \otimes \mathcal{H}_e}_{2N~\text{times}})$} subspace of the Fock space $\mathcal{F}_e$.

\subsection{Localized spin operators}\label{sec:exact}
The matrix representation of the localized spin operator $\hat{\bold{S}}=(\hat{S}^x,\hat{S}^y,\hat{S}^z)$ is given by 
\begin{subequations}\label{eq:spin_mat}
\begin{eqnarray}
\bra{m\pm1}\hat{S}^x\ket{m} &=& \frac{1}{2}\sqrt{S(S+1) - m(m\pm1)}, \label{eq:sx_mat}
\\
\bra{m\pm1}\hat{S}^y\ket{m} &=& \frac{1}{2i} \sqrt{S(S+1) - m(m\pm1)}, \label{eq:sy_mat}
\\
\bra{m}\hat{S}^z\ket{m} &=& m , \label{eq:sz_mat}
\end{eqnarray}
\end{subequations}
where $\ket{m}$ is an eigenstate of $\hat{S}^z$; $m \in \{-S,-S+1,\cdots,S-1,S\}$; and $\hat{S}^\alpha$ is a $(2S+1)\times(2S+1)$ matrix acting in the single-site subspace $\mathcal{H}_i$ of $\mathcal{H}_\mathrm{lspins}$. For the chain in Fig.~\ref{fig:fig1}(a) composed of $N$ sites hosting spin-$S$ localized spins, their operators act in the total space of all localized spins $\mathcal{H}_\mathrm{lspins}$ [Eq.~\eqref{eq:hlspins}] as 
\begin{equation}\label{eq:spinop_N}
\hat{S}^\alpha_i = \underbrace{\id \otimes \id \cdots \id \otimes \id}_{i-1~{\rm times}} \otimes~\hat{S}^\alpha \otimes \underbrace{\id \otimes \id \cdots \id \otimes \id}_{N-i~{\rm times}},
\end{equation}
where $\id$ is  $(2S+1)\times(2S+1)$ unit matrix. 

\subsection{Truncated Holstein-Primakoff transformation}\label{sec:hptruncated}
The HP transformation shown in Eq.~\eqref{eq:hp} expresses localized spin operators in terms of bosonic operators. However, to make MBPT for such bosonic operators tractable~\cite{Harris1971,Hamer1992}, one typically expands the square root of Eq.~\eqref{eq:hp} in a power series in $x = \hat{n}_i/2S$
\begin{equation}\label{eq:tay}
(1 - x)^{1/2} = \sum_{n=0}^{\infty} \frac{2n!}{4^n(1-2n)n!^2}x^n \approx \sum_{n=0}^{N_T} \frac{2n!}{4^n(1-2n)n!^2}x^n,
\end{equation}
which is further truncated~\cite{Chudnovsky2006,Tupitsyn2008,Yuan2020,Takei2019,Mook2021,Elyasi2020} to a finite number of terms $N_T$. Inserting this result in Eq.~\eqref{eq:hp}, and using thus obtained $\hat{S}^\alpha_i$ in Eq.~\eqref{eq:spin_ham}, we can re-write
\begin{equation}\label{eq:split_ham}
\hat{H}_\mathrm{lspins} = \hat{H}_{0} + \hat{H}_\mathrm{int},
\end{equation}
as the sum of two terms. Here 
\begin{equation}\label{eq:ni_ham}
\hat{H}_0 = -J_H (N-1)S^2 + 2J_H S \sum_{i=1}^{N} \hat{n}_i - J_H S \sum_{\braket{ij}} (\hat{a}_i^\dagger a_j + a_ia_j^\dagger), 
\end{equation}
is one-particle Hamiltonian of noninteracting HP bosons covered in textbook literature~\cite{Mahan2011,Chudnovsky2006}, whereas  
\begin{eqnarray}\label{eq:mag_int_ham}
\hat{H}_\mathrm{int} = -J_H \sum_{\braket{ij}} &\bigg[&
  \hat{n}_i\hat{n}_j
- \frac{\hat{a}_i\hat{a}_j^\dagger \hat{n}_j}{4S} 
- \frac{a_i^\dagger \hat{n}_j \hat{a}_j}{4S} 
- \frac{\hat{n}_i\hat{a}_i \hat{a}_j^\dagger}{4S} \nonumber \\ 
&-& \frac{\hat{a}_i^\dagger\hat{n}_i\hat{a}_j}{4S} 
+ \frac{\hat{n}_i\hat{a}_i\hat{a}_j^\dagger \hat{n}_j}{16S^2}
+ \frac{\hat{a}_i^\dagger \hat{n}_i\hat{n}_j\hat{a}_j}{16S^2} 
\bigg] + \cdots, \nonumber \\
\end{eqnarray}
is composed of many-particle interacting terms that we write explicitly for truncation number $N_T=1$ to emphasize how nontrivial multi-boson interactions 
arise even in this lowest order truncated HP transformation. The bosonic operator $\hat{a}^\dagger$ is represented by an infinite matrix
\begin{equation}\label{eq:adag_mat}
\hat{a}^\dagger = \begin{pmatrix} 
0 & 0& 0  & \cdots &0& \cdots \\
\sqrt{1} & 0  & 0 & \cdots &0 &\cdots \\
0 & \sqrt{2}  & 0 & \cdots &0& \cdots \\
0 & 0 & \sqrt{3}  & \ddots& \vdots & \cdots \\
0 & 0 & 0 & \cdots& \sqrt{n} & \cdots \\
\vdots & \vdots & \vdots & \vdots& \vdots & \ddots
\end{pmatrix},
\end{equation}
for a single site, so that matrix representation of $\hat{a}^\dagger_i$ in the case of $N$ sites is given by
\begin{equation}\label{eq:adag_N}
\hat{a}^\dagger_i = \underbrace{\id \otimes \id \cdots \id \otimes \id}_{i-1~{\rm times}} \otimes~\hat{a}^\dagger \otimes \underbrace{\id \otimes \id \cdots \id \otimes \id}_{N-i~{\rm times}},
\end{equation}
where $\id$ is the unit matrix of the same size as $\hat{a}^\dagger$. The matrix representation of operator $\hat{a}_i$ is the Hermitian conjugate of $\hat{a}^\dagger_i$. 

\subsection{Resummed Holstein-Primakoff transformation}\label{sec:hpresum}
In numerical calculations,  $\hat{a}^\dagger$ or $\hat{a}$ are first truncated to a finite  $N_B \times N_B$ matrices, so that the matrix representation of localized spin operators is then composed of matrix blocks associated with physical states $\left\{ \ket{n} \right\}_{n=0, \cdots, 2S}$ and unphysical states $\left\{ \ket{n}\right\}_{n=2S+1, \cdots, N_B}$
\begin{equation}\label{eq:phys_states}
    \hat{S}^{\pm, z} = \begin{pmatrix}
    \hat{S}^{\pm, z}_\mathrm{phys} & \Delta_\mathrm{c} \\
    \Delta^\dagger_\mathrm{c} & \hat{S}^{\pm, z}_\mathrm{unphys}
    \end{pmatrix}, 
\end{equation}
where $\Delta_\mathrm{c}$ is the coupling between physical and unphysical states. The numerically exact computation of the square root of an operator in Eq.~\eqref{eq:hp} ensures $\Delta_\mathrm{c} = 0$, but Taylor expansion of square root in Eq.~\eqref{eq:tay} leads to $\Delta_\mathrm{c} \neq 0$ which, therefore, couples the physical and unphysical states. This feature reveals the trouble with the truncated HP transformation.

Alternatively, Refs.~\cite{Vogl2020, Konig2021} have recently proposed a \emph{resummed HP transformation} that furnishes a polynomial expansion for the square root in Eq.~\eqref{eq:hp}
\begin{equation}\label{eq:hp_resum}
    \hat{S}^+_i 
     \approx \sqrt{2S} \left[ \sum_{n=0}^{N_\mathrm{max}}C_n (\hat{a}^\dagger_i)^n(\hat{a}_i)^n\right] \hat{a}_i,
\end{equation}
where the iterative relation for coefficients $C_n$
\begin{equation}\label{eq:hp_resum_qn}
C_n = \frac{1}{n!}\bigg( 1 - \frac{n}{2S}\bigg)^{1/2} - \sum_{m=0}^{n - 1} \frac{C_m}{(n-m)!}, 
\end{equation}
was derived in Ref.~\cite{Vogl2020} by using flow-equations, whereas an equivalent closed-form expression
\begin{equation}\label{eq:hp_resum_2nd}
C_n = \sum_{k=0}^n (-1)^{n-k} \frac{n!}{k!\,(n-k)! }\bigg( 1 - \frac{k}{2S}\bigg)^{1/2},
\end{equation}
was derived  in Ref.~\cite{Konig2021}  by using Newton-series expansion. Equation~\eqref{eq:hp_resum} ensures that for $N_\mathrm{max} = 2S$ the matrix-block $S^{\pm, z}_\mathrm{phys}$ associated with the physical states is \emph{exact}, whereas coupling between the physical and unphysical states is $\Delta_\mathrm{c} = 0$, which makes nonzero submatrix $\hat{S}^{\pm, z}_\mathrm{unphys}$ irrelevant for all practical purposes.

\subsection{Relationship between localized spin operators and their mapping to Holstein-Primakoff bosons}\label{sec:exacthp}
For physically transparent understanding of  the relationship between localized spin operators and their mapping to HP bosons, let us consider an example of 1D chain of $N=7$ sites hosting spin-$\frac{5}{2}$ localized spins. We use arrows of different length
\begin{equation}
    \bigg\downarrow, \big\downarrow, \downarrow, \uparrow, \big\uparrow, \bigg\uparrow,
\end{equation}
to denote eigenvalues $S^z_i$ of localized spin operator $\hat{S}^z_i$ [Eq.~\eqref{eq:sz_mat}] with $m=-5/2$, $-3/2$, $-1/2$, $1/2$, $3/2$, $5/2$, respectively, as illustrated in Fig.~\ref{fig:fig1}(c). The ferromagnetic ground state of this system 
\begin{equation}\label{eq:hp_vac_state}
\bigg\vert\bigg\uparrow~\bigg\uparrow~\bigg\uparrow~\bigg\uparrow~\bigg\uparrow \bigg\rangle \equiv \ket{0}
\end{equation}
is identical to HP bosonic vacuum state~$\ket{0}$ with zero HP bosons on each site $n_i=0$ and, therefore,  total number of HP bosons 
\begin{equation}\label{eq:nmag}
N_\mathrm{mag} = \sum_i n_i, 
\end{equation}
also being zero, $N_\mathrm{mag}=0$. Inside the ket vector on the left hand side (LHS) of  Eq.~\eqref{eq:hp_vac_state}, we indicate eigenstate with eigenvalue $m=5/2$ of the localized spin operator $\hat{S}^z_i$ for all sites $i=1$ to $i=4$. Equation~\eqref{eq:hp_vac_state} is proved by noting that $\ket{0}$ on the right hand side (RHS) of Eq.~\eqref{eq:hp_vac_state} and ket on the LHS of Eq.~\eqref{eq:hp_vac_state}  are both eigenstates of the same operator $\hat{S}^z_i$ with eigenvalue $m=5/2$ i.e.,  
\begin{subequations}\label{eq:hp_vac}
\begin{eqnarray}
    \hat{S}^z_i \bigg\vert\bigg\uparrow~\bigg\uparrow~\bigg\uparrow~\bigg\uparrow~\bigg\uparrow \bigg\rangle
    &=& \frac{5}{2} \bigg\vert\bigg\uparrow~\bigg\uparrow~\bigg\uparrow~\bigg\uparrow~\bigg\uparrow \bigg\rangle, \\
    \hat{S}^z_i\ket{0} = (5/2 - \hat{n}_i)\ket{0}
    &=&  5/2\ket{0},
\end{eqnarray}
\end{subequations}
so they must be identical. Thus, creating $n_i=1$ or $n_i = 2$ HP bosons on site $i=1$, which we depict  by
\begin{eqnarray}\label{eq:hp_one_mag}
\hat{a}_1^\dagger \ket{0} 
&=& \bigg\vert\big\uparrow~\bigg\uparrow~\bigg\uparrow~\bigg\uparrow~ \bigg\uparrow \bigg\rangle, \\ \label{eq:hp_two_mag}
(\hat{a}_1^\dagger)^2 \ket{0} &=& \bigg\vert \hspace{-0.04cm}\uparrow~ \bigg\uparrow~\bigg\uparrow~\bigg\uparrow~\bigg\uparrow \bigg\rangle,
\end{eqnarray}
respectively, corresponds to reducing the size of localized spin on site $i=1$ by $1$ or $2$ units $\hbar$, i.e.,  $m=5/2 \mapsto 3/2$ in Eq.~\eqref{eq:hp_one_mag} and $m=5/2 \mapsto 1/2$ in Eq.~\eqref{eq:hp_two_mag}. Similarly, the state with a total of $N_\mathrm{mag} = 2$ HP bosons created on 
different sites $i=1$ and $i=3$ is depicted by
\begin{equation}\label{eq:hp_two_mag2}
\hat{a}^\dagger_1 \hat{a}^\dagger_3 \ket{0} =  \bigg\vert\big\uparrow~ \bigg\uparrow~\big\uparrow~\bigg\uparrow~\bigg\uparrow~\bigg\rangle.
\end{equation}
Thus, creating a total of $N_\mathrm{mag} \in 0, 1, 2, \ldots$ [Eq.~\eqref{eq:nmag}] HP bosons is interpreted physically as the reduction of the total localized $z$-spin by $N_\mathrm{mag}$ units. Since in quantum state $(\hat{a}_i^\dagger)^n|0\rangle$ the expectation value of the $z$-component of localized spin operator is $\braket{\hat{S}^z_i} = S - n$, the constraint $0 \le n \le 2S$ (i.e., at a given site $i$ one {\em cannot} create more than $2S$ HP bosons) must be obeyed in order to remain in the subspace of physical states [Eq.~\eqref{eq:phys_states}].

\subsection{Numerically exact time evolution of quantum many-body states}\label{sec:cn}

The solution of time-dependent Schr\"{o}dinger equation for quantum many-body state $\ket{\Psi(t)}$
\begin{equation}\label{eq:psievolve}
	i\hbar \frac{d\ket{\Psi(t)}}{dt} = \hat{H}(t) \ket{\Psi(t)},
\end{equation}
is formally given by
\begin{equation}\label{eq:evolutionop}
	\ket{\Psi(t+\delta t)} = \mathcal{T} \exp \left( -\frac{i}{\hbar}\int\limits_t^{t+\delta t}dt' \hat{H}(t') \right) \ket{\Psi(t)},
\end{equation}
where $\mathcal{T}$ is the time-ordering operator. While many numerical algorithms are available to propagate Eq.~\eqref{eq:evolutionop}, including direct computation of matrix exponential when $\hat{H}$ is time independent, in general by using sufficiently small $\delta t$ and by considering $\hat{H}(t)$ to be constant over such $\delta t$ the Crank-Nicolson algorithm 
\begin{equation}\label{eq:cn}
	\left( 1 + \frac{i\delta t}{2\hbar}\hat{H}(t) \right) \ket{\Psi(t+\delta t)} = \left( 1 - \frac{i\delta t}{2\hbar}\hat{H}(t) \right) \ket{\Psi(t)},
\end{equation}
we employ offers propagation scheme that is unitary, accurate to second order in $\delta t$, and unconditionally stable~\cite{Wells2019}. 

Using thus obtained $\ket{\Psi(t)}$, the time evolution of the expectation value of the $\alpha$-component of  localized spin operator on site $i$ is given by 
\begin{equation}\label{eq:expectationspin}
	\langle \hat{S}^\alpha_i \rangle (t) = \bra{\Psi(t)}\hat{S}^\alpha_i\ket{\Psi(t)}.
\end{equation} 
When localized spin operators are represented directly by finite size matrices in Eqs.~\eqref{eq:spin_mat} and ~\eqref{eq:spinop_N}, the corresponding expectation values $\langle S^\alpha_i \rangle$ are ~\emph{numerically~exact} and, therefore, serve as a benchmark for alternative computation of the same expectation value when $\hat{S}^\alpha_i$ are represented by polynomial expressions in bosonic operators introduced in Secs.~\ref{sec:hptruncated} and ~\ref{sec:hpresum}.

\subsection{From Holstein-Primakoff bosons to one- or two-magnon Fock states}\label{sec:magnons}

In contrast to HP bosons created on a given site, $\hat{a}_i^\dagger|0\rangle$, which are not the eigenstates of $\hat{H}_0$ in Eq.~\eqref{eq:ni_ham}, one-magnon states are linear combinations of $\hat{a}_i^\dagger|0\rangle$ which diagonalize Hamiltonian $\hat{H}_0$ (but with periodic boundary conditions included)
\begin{equation}\label{eq:eigenmagnon}
	\hat{H}_0 |q\rangle = [E_0 + \hbar \omega(q)]|q\rangle.
\end{equation}
Thus, they can be visualized  bosonic quasiparticle which carries momentum $\hbar q$ (assuming 1D chains we use as  examples) and angular momentum $\hbar$ and is completely ``delocalized'' over all sites. Here $E_0=-2J_H S^2 N$ is the ground state energy of a ferromagnetic spin chain. 

To find explicit expression for excited eigenstate $|q\rangle$, we consider 1D chain [Fig.~\ref{fig:fig1}(a)] composed of $N$ sites each of which is hosting spin-$1$ localized spin and with periodic boundary conditions so that its first and last site are coupled by $J_H$ in $\hat{H}_0$ in Eq.~\eqref{eq:ni_ham}. For the clarity  of notation, we use $\downarrow$, $\odot$, and $\uparrow$, to denote eigenstates of localized spin operator $\hat{S}^z_i$ with eigenvalues [Eq.~\eqref{eq:sz_mat}] $m=-1$, $m=0$, and $m=1$, respectively. The one-magnon state is then given by 
\begin{equation}\label{eq:locs_one_mag}
    \ket{q} \equiv \frac{1}{\sqrt{N}}
    \sum_{n=0}^{N-1}e^{i q x_n}\ket{ \underbrace{\uparrow \cdots\uparrow}_{n~\text{times}} \odot  \underbrace{\uparrow\cdots \uparrow}_{N-n-1~\text{times}}},
\end{equation}
where $x_n = na_0$ is the real-space position of the localized spin on site $n+1$ and $a_0$ is the lattice spacing. The corresponding magnon energy-momentum dispersion is  $\hbar \omega(q) = 2JS[1-\cos(qa_0)]$. The expectation value of the total $z$-spin operator of localized spins in state $|q\rangle$ is given by
\begin{equation}\label{eq:sam_reduction}
   \bra{q}\sum_{i=1}^N \hat{S}^z_i\ket{q} = (NS - 1),
\end{equation}
%
%
\begin{figure}
	\centering
	\includegraphics[width=\linewidth]{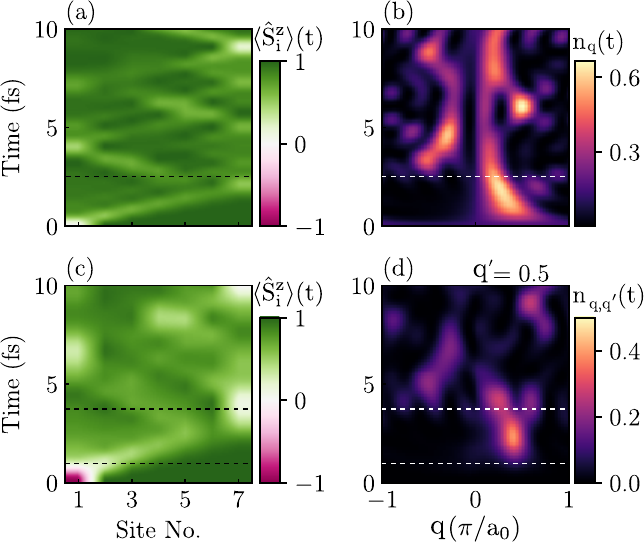}
	\caption{ (a) Spatio-temporal profile of the expectation value $\braket{\hat{S}^z_i}(t)$ across 1D spin chain in Fig.~\ref{fig:fig1}(a) composed of $N=7$ sites hosting spin-$1$ localized spins where at $N_\mathrm{mag} = 1$ HP boson is created  at initial time $t=0$  on site $i=1$, $\ket{\Psi_1(t=0)} = \ket{\odot \uparrow \uparrow\uparrow \uparrow \uparrow \uparrow}$. (b) The corresponding probability, $n_q(t) = |\braket{q|\Psi_1(t)}|^2$, of finding quantum many-body state $\ket{\Psi_1(t)}$ at later times $t>0$ in {\em one-magnon Fock state} $|q \rangle$ of momentum $q$ as  HP boson propagates from the left to the right edge of the chain [``white traces'' in (a)], thereby switching from $q>0$ to $q < 0$  when reflection occurs near the boundary on site $i=7$ and $t \gtrsim 2.5$ fs (indicated by dashed horizontal lines). Panel (c) is counterpart of panel (a) when  $N_\mathrm{mag} = 2$ HP bosons are created at $t=0$ on site $i=1$, $\ket{\Psi_2(t=0)} = \ket{\downarrow \uparrow \uparrow \uparrow \uparrow \uparrow \uparrow}$, with panel (d) showing the corresponding probability, $n_{q, q'}(t) = |\braket{q, q'|\Psi_2(t)}|^2$, of finding quantum many-body state $\ket{\Psi_2(t)}$  at later times $t>0$ in {\em two-magnon Fock state} $|q,q'\rangle$. Dashed horizontal lines in panel (d) mark times \mbox{$t = 1$ fs} and \mbox{$t = 4$ fs}. In all panels we set \mbox{$J_H = 1$ eV}.}
	\label{fig:fig3}
\end{figure}
which indicates that creation of magnon with wavevector $q$ removes one unit of total $z$-spin from the ferromagnetic ground state. {\em Because of this feature, presence of one HP boson or one HP magnon is labeled by the same $N_\mathrm{mag} = 1$ throughout the paper}. In addition, the expectation value of the localized $z$-spin operator at arbitrary site $i$ 
\begin{equation}\label{eq:loc_sam_reduction}
    \bra{q} \hat{S}^z_i \ket{q} = \left( S - \frac{1}{N}\right),
\end{equation}
shows that excitation of one HP magnon reduces the $z$-component of each localized spin by $1/N$. This rigorous quantum-mechanical result justifies the LLG picture~\cite{Kim2010} of spin wave in which classical vectors of localized spins precess with frequency $\omega$ and with some small cone angle around the $z$-axis, while the phase of the precession of adjacent vectors varies harmonically in space over the wavelength $\lambda$.

In second-quantization description produced by HP transformation, $\ket{q} = \hat{b}_q^\dagger\ket{0}$ is {\em one-magnon Fock state}~\cite{Quirion2020} where the creation operator of HP magnon is given by
\begin{eqnarray}\label{eq:fockmagnon}
        \hat{b}^\dagger_q = \frac{1}{\sqrt{N}}\sum_{n=0}^{N-1} e^{ikx_n}\hat{a}^\dagger_{n+1}.
\end{eqnarray}
Note that such one-magnon Fock state has been realized experimentally only very recently in a millimeter-sized ferrimagnetic crystal and detected by superconducting qubit as quantum sensor~\cite{Quirion2020}, thereby representing  a counterpart in quantum magnonics of a single-photon detection from 
quantum optics.

Note that in spintronics and magnonics literature~\cite{Kamra2016} one also finds $\hat{a}_i^\dagger|0\rangle$ denoted as ``one magnon created in real space at position $i$'' while $\hat{b}^\dagger_q |0\rangle$ is ``one magnon created in the reciprocal space with momentum $\hbar q$''. However, the former is not an eigenstate of Hamiltonian in Eq.~\eqref{eq:eigenmagnon}, while the later is, so we differentiate between them by using ``HP boson'' for the former and  ``HP magnon'' for the latter. As already highlighted, for both situations we use label $N_\mathrm{mag} =1$ for simplicity of notation because in both cases one unit of total $z$-spin is removed from the ferromagnetic ground state [Eq.~\eqref{eq:sam_reduction}].

Nevertheless, we illustrate the distinction between HP boson and HP magnon by initializing $N=7$ site chain [Fig.~\ref{fig:fig1}(a)] in quantum state $\ket{\Psi_1(t=0)} = \ket{\odot \uparrow \uparrow \uparrow \uparrow \uparrow \uparrow}$ in Figs.~\ref{fig:fig3}(a) and ~\ref{fig:fig3}(b); or in quantum state \mbox{$\ket{\Psi_2(t=0)}=\ket{\downarrow \uparrow \uparrow \uparrow \uparrow \uparrow \uparrow}$} in Figs.~\ref{fig:fig3}(c) and ~\ref{fig:fig3}(d). This means that $N_\mathrm{mag}=1$ HP boson is created on site $i=1$ at $t=0$ in the former case; while ``full spin flip'' of localized spin on site $i=1$ in the latter case means that $N_\mathrm{mag} = 2$ HP bosons are created on site $i=1$. Besides pedagogical value, such initial states and one or two magnon propagation including magnon bound states, has also been probed experimentally in ultracold atoms in an optical lattice where tracking of the localized spin expectation values is possible with single-spin and single-site resolution~\cite{Fukuhara2013}.

Since $\ket{\Psi_1(t=0)}$ is not an eigenstate, it evolves in time to produce spatio-temporal profile of the expectation value $\langle \hat{S}^z_i \rangle  (t)$ [Fig.~\ref{fig:fig3}(a)] in quantum state $\ket{\Psi_1(t)}$. For quantum time evolution we use the scheme explained in Sec.~\ref{sec:cn} where interacting Hamiltonian $\hat{H}_\mathrm{lspins}$ from Eq.~\eqref{eq:hlspins} is plugged in, but since only one HP boson is excited this is equivalent to using noninteracting $\hat{H}_0$ in Eq.~\eqref{eq:ni_ham}. The ``white trace''  in Fig.~\ref{fig:fig3}(a) visualizes how HP boson moves from the left to the right edge of the chain while undergoing reflection  on site $i=7$ at \mbox{$t \approx 2.5 $ fs}, as indicated by horizontal dashed line, followed by multiple back-and-forth reflections. Note that since 1D chain in Fig.~\ref{fig:fig1}(a) has open boundary conditions, its low-energy excited eigenstates differ~\cite{Haque2010} from textbook~\cite{Mahan2011,Chudnovsky2006}  HP magnons $|q\rangle$ in Eq.~\eqref{eq:locs_one_mag} as eigenstates of interacting localized spin systems with translational invariance (which is, therefore, either infinite or finite length but with periodic boundary conditions). Figure~\ref{fig:fig3}(b) visualizes the  overlap, $n_q(t) = |\braket{q|\Psi_1(t)}|^2$, between many-body quantum state  $\ket{\Psi_1(t)}$  with one HP boson and one-magnon Fock state $|q\rangle$. Large values of $n_q(t)$ are observed in the region where $q>0$ and \mbox{$t \lesssim 2.5$ fs}, coinciding with left-to-right motion of HP boson in Fig.~\ref{fig:fig3}(a), which signifies excitation of $N_\mathrm{mag}=1$ magnon with positive momentum. On the other hand, after reflection of the HP boson at the boundary (i.e., site $i=7$) and \mbox{$t \approx  2.5$ fs}, a rapid rise of $n_{q}(t)$ in the $q < 0$ region is observed which indicates excitation of $N_\mathrm{mag}=1$ magnon with negative momentum. This is consistent with the intuitive picture of HP boson reflecting back-and-forth between the hard walls of our 1D chain with open boundary conditions. 

The Fock states of $N_\mathrm{mag}=2$ magnons carrying momentum $\hbar q$ and $\hbar q'$ are defined by~\cite{Morimae2005}   
\begin{equation}\label{eq:two_mag_bs}
	\ket{q, q'} = \sum_{n>m} f_{mn}(q, q') \ket{\underbrace{\uparrow\cdots\uparrow}_{m-1~\text{times}} \odot \uparrow\cdots\uparrow \odot \underbrace{\uparrow \cdots \uparrow}_{N-n~\text{times}}},
\end{equation}
where 
\begin{equation}
	f_{mn}(q, q') = \frac{1}{\mathcal{N}} \bigg[ e^{i q x_m}e^{iq'x_n} + e^{iq'x_n}e^{iqx_m}\bigg].
\end{equation}
Here $\mathcal{N}$ is the normalization constant, and $f_{mn}(q, q')$ is symmetric under exchange $q \leftrightarrow q'$ ensuring \mbox{$\ket{q, q'} = \ket{q', q}$} in order to satisfy the symmetrization postulate of quantum mechanics for bosonic particles---as manifestly encoded by second-quantization formalism,  $\ket{q, q'} = \hat{b}^\dagger_q\hat{b}^\dagger_{q'}\ket{0} = \hat{b}^\dagger_{q'}\hat{b}^\dagger_{q}\ket{0} = \ket{q', q}$.

Figure~\ref{fig:fig3}(c) plots spatio-temporal profile of $\langle \hat{S}^z_i \rangle (t)$ in quantum state $\ket{\Psi_2(t)}$ starting from \mbox{$\ket{\Psi_2(t=0)} = \ket{\downarrow \uparrow \uparrow \uparrow \uparrow \uparrow \uparrow}$}. For quantum time evolution we use the scheme explained in Sec.~\ref{sec:cn} where interacting Hamiltonian $\hat{H}_\mathrm{lspins}$ from Eq.~\eqref{eq:hlspins} is plugged in, so that two HP bosons are correlated by: (\emph{i}) bosonic statistics; (\emph{ii}) interactions in $\hat{H}_\mathrm{int}$ [Eq.~\eqref{eq:mag_int_ham}] where $N_T \rightarrow \infty$. The two HP bosons propagate immediately from the left to the right for $t>0$, as shown by ``white traces'' in Fig.~\ref{fig:fig3}(c). The corresponding overlap, $n_{q, q'}(t) = |\braket{q, q'|\Psi_2(t)}|^2$, in Fig.~\ref{fig:fig3}(d) is $n_{q, q'}(t) = 0$ for $t \lesssim 1$ fs which is explained by Eq.~\eqref{eq:two_mag_bs} where two-magnon Fock state is composed  of terms containing two HP bosons on different sites $m\neq n$. Since at $t=0$ the two HP bosons are on the same site $i=1$, we find $n_{q, q'}(t=0)=0$. However, this holds until $t \lesssim 1$ fs (indicated by horizontal dashed line), after which the two HP bosons are physically separated in real space, as confirmed by the emergence of nonzero values of $n_{q, q'}(t)$ thereafter. We also note that for $1\lesssim t \lesssim 4$ fs (indicated by horizontal dashed line) the region near \mbox{$q = 0.5$ $\pi/a_0$} shows large values of $n_{q, q'}(t)$, and since \mbox{$q'=0.5$ $\pi/a_0$} is fixed for all values of $q$ in Fig.~\ref{fig:fig3}(d), we can conclude that two HP bosons posses nearly the same velocity. Beyond $t~\approx~4$~fs, nonzero values of $n_{q, q'}(t)$ in the region with $q<0$ and $q>0$ coexist, which indicates that one HP boson moves toward the right while the other moves toward the left edge of the chain.

\subsection{Retarded and lesser one-particle Green functions}\label{ref:gf_sf}
The fundamental quantities of nonequilibrium GF formalism~\cite{Stefanucci2013,Schlunzen2020} for fermions are the one-particle retarded GF

\begin{equation}\label{eq:ele_GF_ret}
G^r_{i\sigma, j\sigma'}(t,t') =
-i\hbar^{-1}\Theta(t-t')\braket{\{\hat{c}_{i\sigma}(t),\hat{c}_{j\sigma'}^\dagger(t') \}}, 
\end{equation}
and the one-particle lesser GF
\begin{equation}\label{eq:ele_GF_less}
G^<_{i\sigma,j\sigma'}(t,t') = 
i\hbar^{-1} \braket{\hat{c}_{j\sigma'}^\dagger(t')\hat{c}_{i\sigma}(t)},
\end{equation}
which describe the density of available quantum states and how electrons occupy those states, respectively. Here $\Theta(t-t')$ is the Heaviside-function; $\hat{c}_{i\sigma}(t)$ indicates Heisenberg picture time evolution of $\hat{c}_{i\sigma}$; and $\braket{\cdots} = \mathrm{Tr}(\hat{\rho} \cdots)$ is the quantum statistical average, where $\hat{\rho}$ is the density operator of the system at $t=0$.  Analogously, the bosonic one-particle retarded GF is defined by 
\begin{equation}\label{eq:mag_GF_ret}
D^r_{ij}(t,t') =
-i\hbar^{-1}\Theta(t-t')\braket{[\hat{a}_{i}(t),\hat{a}_{j}^\dagger(t')]},
\end{equation}
and the lesser GF is defined by
\begin{equation}\label{eq:mag_GF_less}
D^<_{ij}(t,t') = 
-i \hbar^{-1}\braket{\hat{a}_{j}^\dagger(t')\hat{a}_{i}(t)}. 
\end{equation}
In equilibrium or in steady-state nonequilibrium, these GFs  depend solely on $\tau = t-t'$ and can be Fourier transformed to energy domain~\cite{Mahfouzi2014}, such as 
\begin{equation}\label{eq:ele_FTGF}
G^{r,<}_{i\sigma,j\sigma'}(E) = 
\int\displaylimits_{-\infty}^{+\infty} d\tau \,
G^{r,<}_{i\sigma,j\sigma'}(\tau) e^{iE\tau/\hbar},
\end{equation}
for electrons; and 
\begin{equation}\label{eq:mag_FTGF}
D^{r,<}_{ij}(E) = \int\displaylimits_{-\infty}^{+\infty} d\tau 
D^{r,<}_{ij}(\tau) e^{iE\tau/\hbar},
\end{equation}
for bosons.

\subsection{Spectral function for electrons and magnons}
The  electronic spectral function $A(E)$, or the ``interacting density of states''~\cite{Balzer2011,Nocera2018}, is computed using the retarded GF in Eq.~\eqref{eq:ele_FTGF} as
\begin{eqnarray}\label{eq:ele_spec_f}
A(E) &=&
-2\sum_{i=1}^{N}\sum_{\sigma=\uparrow, \downarrow} \mathrm{Img}[G^r_{i\sigma,i\sigma}(E)] \nonumber \\
&=& \sum_{k} W_k^+ \delta(E-\Delta_k) + W_k^-\delta(E+\Delta_k),
\end{eqnarray}
where $\Delta_k = (E_k - E_0)$ and $E_k$ are the eigenenergies of quantum many-body Hamiltonian $\hat{H}$ [Eq.~\eqref{eq:ham_sys}]. The prefactors of $\delta$-function  in $A(E)$  
\begin{subequations}\label{eq:ele_weight}
\begin{eqnarray}
W_k^+ &=& \sum_{i=1}^N\sum_{\sigma=\uparrow, \downarrow}|\bra{\Psi_k} \hat{c}_{i\sigma}^\dagger\ket{\Psi_0}|^2, \\
W_k^- &=& \sum_{i=1}^N\sum_{\sigma=\uparrow, \downarrow}|\bra{\Psi_k} \hat{c}_{i\sigma}\ket{\Psi_0}|^2,
\end{eqnarray}
\end{subequations}
define the ``weight'' of the many-body eigenstate $\ket{\Psi_k}$ within $A(E)$. Since the ground state $\ket{\Psi_0}$ is an eigenstate of the electron number operator $\hat{N}_e$ [Eq.~\eqref{eq:ele_num}], it has a well-defined number  of electrons $N_e$. Thus, the action of $\hat{c}_{i\sigma}^\dagger$ and $\hat{c}_{i\sigma}$ on $\ket{\Psi_0}$ in Eq.~\eqref{eq:ele_weight} reveals that the $\delta$-function peaks at $E = \pm \Delta_k$ can only be contributed by those quantum many-body eigenstates $\ket{\Psi_k}$ which describe systems containing $N_e\pm 1$ electrons. 

Similarly, the bosonic spectral function $D(E)$ is evaluated using the bosonic retarded GF  in Eq.~\eqref{eq:mag_FTGF}
\begin{eqnarray}\label{eq:mag_spec_f}
D(E) &=& -2\sum_{i=1}^{N}\mathrm{Img}[D^r_{ii}(E)] \nonumber \\
&=& \sum_{k} Q_k^+ \delta(E-\Delta_k) - Q_k^-\delta(E+\Delta_k),
\end{eqnarray}
where 
\begin{subequations}\label{eq:mag_weight}
\begin{eqnarray}
Q_k^+ &=& \sum_{i}|\bra{\Psi_k} \hat{a}_{i}^\dagger\ket{\Psi_0}|^2,
\\
Q_k^- &=& \sum_{i}|\bra{\Psi_k} \hat{a}_{i}\ket{\Psi_0}|^2.
\end{eqnarray}
\end{subequations}
define the ``weight'' of many-body eigenstate $\ket{\Psi_k}$ within $D(E)$. The $\delta$-function peaks in Eq.~\eqref{eq:mag_spec_f} at $E = \pm \Delta_k$ come from many-body eigenstates $\ket{\Psi_k}$. However, unlike the electronic case, they do not have a  well-defined total magnon number $N_\mathrm{mag}$ as they are not eigenstates of the total magnon number operator $\hat{N}_\mathrm{mag} = \sum_{i=1}^N \hat{n}_i$. This is illustrated by Fig.~\ref{fig:fig9}(e) with the structure of one selected many-body eigenstate $\ket{\Psi_k}$ which is a linear combination of many-body states with total magnon number \mbox{$N_\mathrm{mag} = 0$}, \mbox{$N_\mathrm{mag} = 1$} and \mbox{$N_\mathrm{mag} = 2$}. 

Both $A(E)$ and $D(E)$ must satisfy the {\em sum rule} 
\begin{subequations}
    \begin{eqnarray}
        \int\displaylimits_{-\infty}^{+\infty} \frac{dE}{2\pi} A(E)   &=& 2N, \label{eq:ele_sum_rule}\\ 
        \int\displaylimits_{-\infty}^{+\infty} \frac{dE}{2\pi} D(E)  &=& N. 
        \label{eq:mag_sum_rule}
    \end{eqnarray}
\end{subequations}
This feature allows for  physical interpretation where $A(E)dE/2N$ or $D(E)dE/N$ can be viewed as probabilities to find fermion or boson within energy window $dE$ around $E$ in a general quantum many-body system where fermions interact with other fermions and bosons interact with other bosons, as well as with each other. Note that our fermion-boson interacting system, as illustrated in Fig.~\ref{fig:fig1}(b) and described by Hamiltonian in Eq.~\eqref{eq:ham_sys}, includes HP bosons interacting [Eq~\eqref{eq:mag_int_ham}] with other HP bosons when $N_\mathrm{mag} > 1$ and electrons interacting with HP bosons while electron-electron interactions are excluded. Since the sum rule is an exact result, in practical GF calculations it can be employed to test the quality of a verity of analytical and numerical approximations schemes~\cite{Mahfouzi2014}. 
\section{Results and Discussion}\label{sec:randd}
\subsection{Range of validity of truncated HP transformation for nonequilibrium interacting system of magnons}\label{sec:validity_mm}
\begin{figure}
	\centering
	\includegraphics[width=\linewidth]{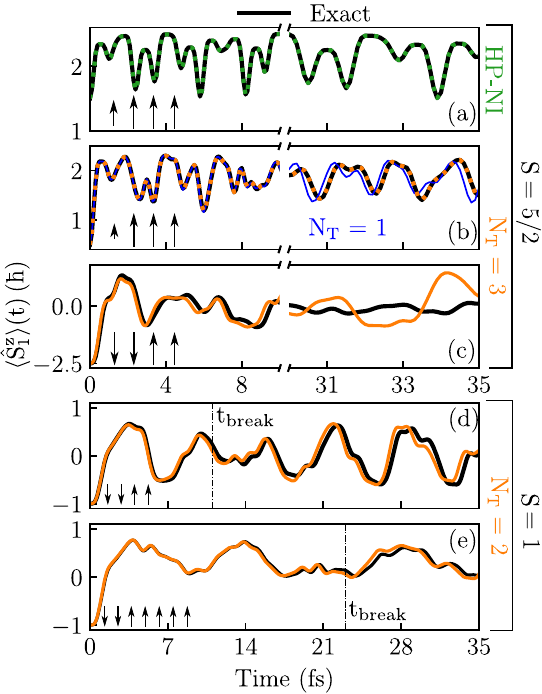}
	\caption{Comparison of the {\em exact} time-dependence $\braket{\hat{S}^z_i}(t)$ (black lines) obtained by using localized spin operators vs.  {\em approximative} time-dependence obtained by using truncated HP transformation [Sec.~\ref{sec:hptruncated}], with a truncation number $N_T$, for 1D quantum spin chain in Fig.~\ref{fig:fig1}(a). The chain is composed of $N=4$ sites hosting spin-$\frac{5}{2}$ [panels (a)--(c)] or spin-$1$ [panels (d)--(e)] localized spins. At $t=0$, $N_\mathrm{mag}$ HP bosons are created on site $i=1$ or both sites $i=1$ and $i=2$, as illustrated by the reduced size of arrows  or their full reversal in the inset at the lower left corner within each panel (see Sec.~\ref{sec:exacthp} for proper association of quantum states to illustration in the insets). In panel~(a), $N_\mathrm{mag} = 1$ so that $\braket{\hat{S}^z_i}(t)$  evaluated (green dotted line) from noninteracting HP boson Hamiltonian [Eq.~\eqref{eq:ni_ham}] is identical to the exact  $\braket{\hat{S}^z_i}(t)$. In panel~(b), $N_\mathrm{mag}=2$ and $\braket{\hat{S}^z_i}(t)$ evaluated from truncated HP transformation with $N_T=1$ (blue line) disagrees with the exact $\braket{\hat{S}^z_i}(t)$, but increasing to $N_T=3$ (orange dotted line) matches the exact result. Nevertheless, in panel~(c) creation of $N_\mathrm{mag} = 10$ HP magnons, by full reversal of two localized spins, while keeping $N_T=3$ leads to disagreement between truncated HP transformation (orange line) and exact (black line) results for $\braket{\hat{S}^z_i}(t)$. In panels (d) and (e), we use chains of $N=4$ and $7$ sites, respectively, where orange lines indicate $\braket{\hat{S}^z_i}(t)$ computed from truncated HP transformation with $N_T=2$. The vertical dot-dash lines in panels (d) and (e) explicitly mark breakdown-time $t=t_\mathrm{break}$ [see also Fig.~\ref{fig:fig5}] at which truncated HP transformation starts to deviate from the exact result for $\braket{\hat{S}^z_i}(t)$. In all panels we set \mbox{$J_H = 1$ eV}.}
	\label{fig:fig4}
\end{figure}
\begin{figure}
	\centering
	\includegraphics[width=\linewidth]{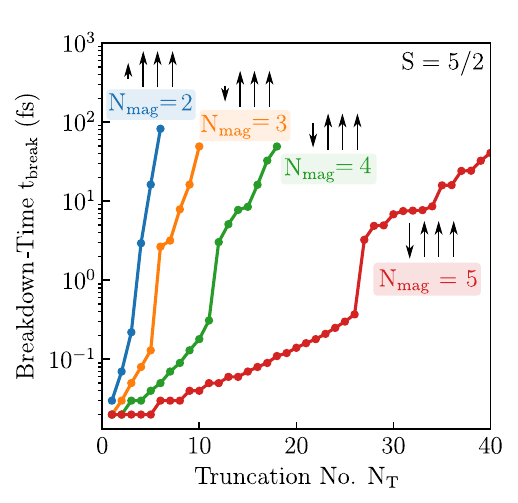}
	\caption{Breakdown-time $t_\mathrm{break}$, as the time at which truncated HP transformation  starts to deviate [see  Figs.~\ref{fig:fig4}(d) and ~\ref{fig:fig4}(e)] from the exact result for $\langle \hat{S}_1^z \rangle (t)$, as function of truncation number $N_T$. The inset near each line indicates the number of HP bosons $N_\mathrm{mag}$ created at $t=0$ within 1D quantum spin chain [Fig.~\ref{fig:fig1}(a)] composed  of $N=4$ sites hosting \mbox{spin-$\frac{5}{2}$} localized spins interacting via Heisenberg exchange \mbox{$J_H = 1$ eV}.}
	\label{fig:fig5}
\end{figure}
Figure~\ref{fig:fig4} compares $\langle \hat{S}_1^z \rangle (t)$ for 1D chain [Fig.~\ref{fig:fig1}(a)], hosting $S=5/2$ or $S=1$ localized spins in the absence of  electrons (i.e., $N_e=0$), computed using the original localized spin operators vs. their mapping to bosonic operators via the  truncated HP transformation. In the ferromagnetic ground state $\ket{\Psi_0}$, the expectation value $\langle \hat{S}^z_i \rangle(t=0) = \braket{\Psi_0|\hat{S}^z_i|\Psi_0} = 5/2$ for all sites $i$ at $t=0$ in Fig.~\ref{fig:fig4}(a)--(c). To initiate nonequilibrium dynamics for times $t>0$, we choose an initial state $\ket{\Psi(0)}$ such that the expectation value of the localized spin on site $i=1$ is reduced by $N_\mathrm{mag}$ units, $\langle \hat{S}^z_{i=1}\rangle(t=0) = \braket{\Psi(0)|\hat{S}^z_{i=1}|\Psi(0)} = (5/2 - N_\mathrm{mag})$, while on other sites it remains  $S^z_{i\neq 1}(t=0) = 5/2$.  This is equivalent to introducing $N_\mathrm{mag}$ HP bosons on site $i=1$ at $t=0$, so that the initial quantum many-body state of HP bosons is given by
\begin{equation}
	\ket{\Psi(0)} = (\hat{a}_1^\dagger)^{N_\mathrm{mag}} \ket{0}.
\end{equation}
When $N_\mathrm{mag} = 1$, Fig.~\ref{fig:fig4}(a) shows that $\langle \hat{S}^z_1 \rangle (t)$, evaluated by truncated HP transformation (green dashed line) solely containing single-particle Hamiltonian $\hat{H}_0$ of noninteracting HP bosons in Eq.~\eqref{eq:ni_ham}, accurately tracks the exact time dependence  (black lines) of $S^z_1(t)$ evaluated using the localized spin operators, as trivially expected. That is, because there is only \emph{one} HP boson in the system,  magnon-magnon interaction terms active within $\hat{H}_\mathrm{int}$ part of the Hamiltonian [Eq.~\eqref{eq:mag_int_ham}] cannot influence the dynamics of localized spins.

To understand the significance of magnon-magnon interaction terms within $\hat{H}_\mathrm{int}$ on the dynamics of localized spins,  we next introduce $N_\mathrm{mag} = 2$ HP bosons on site $i=1$. The time dependence of $\langle \hat{S}^z_1\rangle (t)$ in Fig.~\ref{fig:fig4}(b), evaluated via the truncated HP transformation with truncation number $N_T = 1$ (Sec.~\ref{sec:hptruncated}), matches the exact time evolution obtained using the original localized spin operators  only for short enough times ($0 < t < 10$~fs). At longer times ($30 <t < 35$~fs), discrepancy emerges due to missing effects from $N_T>1$ magnon-magnon interaction terms within $\hat{H}_\mathrm{int}$. Thus, to recover the agreement between two types of calculations at longer times requires increasing $N_T$, such as by using $N_T=3$ (orange dashed lines) in Fig.~\ref{fig:fig4}(b). 
\begin{figure*}
	\centering
	\includegraphics[width=\linewidth]{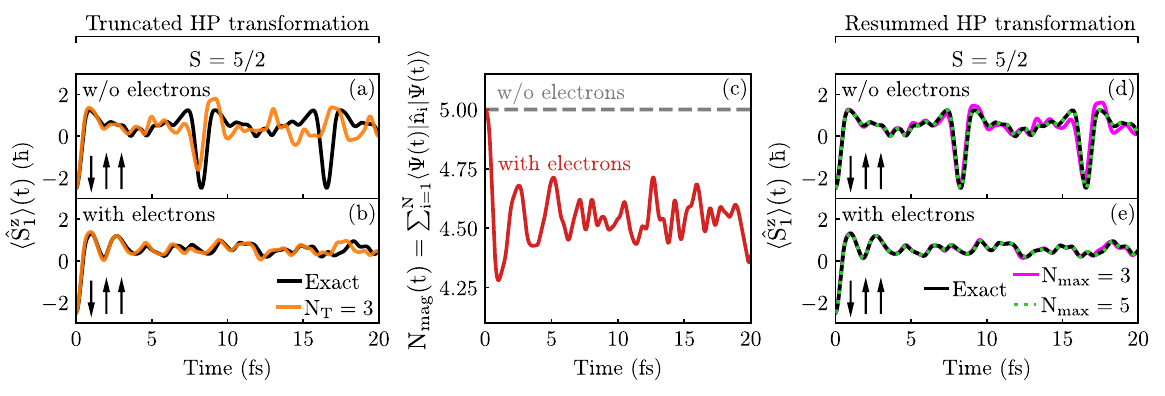}
	\caption{(a) Comparison of the {\em exact} time-dependence $\braket{\hat{S}^z_i}(t)$ (black line) evaluated using localized spin operators vs.  time-dependence computed from truncated HP transformation [Sec.~\ref{sec:hptruncated}], with a truncation number $N_T$ (orange line), for 1D quantum many-body system [Fig.~\ref{fig:fig1}(b)] comprised of $N=3$ sites hosting spin-$\frac{5}{2}$ localized spins which interact with conduction electrons. The electron--localized-spin $sd$ exchange interaction is turned off (\mbox{$J_\mathrm{sd} = 0$ eV}) in (a) for reference, and turned on in (b) using \mbox{$J_\mathrm{sd} = 1$ eV}. The insets in both panels depict the state of localized spins at $t=0$, where the localized spin on site $i=1$ is completely reversed [i.e., $N_\mathrm{mag} = 5$ HP bosons are created on site $i=1$ via Eq.~\eqref{eq:fivehpbosons}] to initiate nonequilibrium dynamics.  When $sd$ interaction is turned on, the disagreement between two types of calculations is actually alleviated when moving from panel (a) to panel (b), which is explained by panel (c) as being  due to a rapid loss of total number of magnons, $N_\mathrm{mag}(t) = \sum_{i=1}^N \bra{\Psi(t)} \hat{n}_i \ket{\Psi(t)}$, from the localized spin subsystem to the electronic subsystem. Panels (d) and (e) are counterparts of panels (a) and (b), respectively, for $\braket{\hat{S}^z_i}(t)$ evaluated using the resummed HP transformation in Eq.~\eqref{eq:hp_resum}, which is {\em approximate} for $N_\mathrm{max} = 3$ but it becomes \emph{exact} for $N_\mathrm{max} = 2S = 5$.}
	\label{fig:fig6}
\end{figure*}
However, progressively larger $N_T$ must be employed (Fig.~\ref{fig:fig5}) to increase the ``breakdown-time'' $t_\mathrm{break}$ [marked in Fig.~\ref{fig:fig4}(d),(e)] at which disagreement between two types of calculations emerges. We define $t_\mathrm{break}$ as the time when the deviation $\Delta~=~\langle \hat{S}^z_1\rangle (t)|_\mathrm{HP} - \langle \hat{S}^z_1\rangle(t)|_\mathrm{exact}$, between $\langle \hat{S}^z_1\rangle(t)|_\mathrm{HP}$ (evaluated from truncated HP transformation  with a truncation number $N_T$) and the exact $\langle \hat{S}^z_1\rangle(t)|_\mathrm{exact}$ becomes larger than the chosen tolerance $|\Delta| > 10^{-4}$. As demonstrated by Figs.~\ref{fig:fig4}(d),(e) and Fig.~\ref{fig:fig5}, $t_\mathrm{break}$ sensitively depends on the density of HP bosons, $N_\mathrm{mag}/N$, whose increase makes magnon-magnon interaction terms within $\hat{H}_\mathrm{int}$ more relevant, thereby requiring larger $N_T$ in Fig.~\ref{fig:fig5}. Figure~\ref{fig:fig4}(d),(e) explicitly confirms this conclusion by showing the effect of  reduced density of HP bosons on the range of validity of truncated HP transformation, where we employ spin-$1$ localized spins allowing us to exactly diagonalize larger chains [than those composed of $N=4$ sites in Fig.~\ref{fig:fig4}(a)--(c) with spin-$\frac{5}{2}$ on each site]. In Fig.~\ref{fig:fig4}(d), at $t=0$ we flip the localized spins on sites $i=1$ and $i=2$ [see inset in Fig.~\ref{fig:fig4}(d)], thereby introducing two HP bosons on each of these sites. Thus, the total number of HP bosons within the system in Fig.~\ref{fig:fig4}(d) is $N_\mathrm{mag} = 4$, whereas the HP boson density is $N_\mathrm{mag}/N = 1$. For such parameters, \mbox{$t_\mathrm{break} \approx 11 $ fs} ($|\Delta| = 0.05$ is chosen solely for visualization of $t_\mathrm{break}$ at fs time scales).  On the other hand, in Fig.~\ref{fig:fig4}(e), where HP boson density is reduced to $N_\mathrm{mag}/N = 0.57$ by making 1D chain longer from $N= 4$ sites to $N=7$ sites, we find that $t_\mathrm{break}$ for truncated HP transformation increases to \mbox{$t_\mathrm{break} \approx 23$ fs}. This observation is easily explained since in longer 1D chains the probability for magnon-magnon scattering events is reduced, which makes inclusion of high-order magnon-magnon interaction terms less important and thus the breakdown-time for truncated HP transformation increases.

Figure~\ref{fig:fig5} demonstrates how for a given breakdown-time \mbox{$t=t_\mathrm{break}$}, the horizontal distance between consecutive curves from left to right increases nonlinearly. This means that $N_T$ needed to accurately track $\langle \hat{S}_1^z \rangle (t)$ via the truncated HP transformation increases nonlinearly with the number of HP bosons $N_\mathrm{mag}$ excited in the system. On the other hand, if we consider the roughly constant slope `$p$' of each curve in Fig.~\ref{fig:fig5} (for the part before a sudden jump), and note the logarithmic scale for the ordinate axis of Fig.~\ref{fig:fig5}, we can conclude that \mbox{$t_\mathrm{break} \propto \exp({pN_T})$}. At first sight, the exponential dependence of $t_\mathrm{break}$ on $N_T$ appears to be favorable i.e., by using larger values of $N_T$ (and hence including more and more multi-magnon terms), we can increase $t_\mathrm{break}$ exponentially and yield accurate dynamics for longer times. However, to obtain a practically tractable MBPT for electron-boson interacting systems~\cite{Marini2018} a small $N_T$ is required but Fig.~\ref{fig:fig5} shows that using small \mbox{$N_T=1$--$5$} allows one to track dynamics of localized spins only up to time \mbox{$t_\mathrm{break} \approx 1.5\hbar/ J_H \approx 15.0$ fs} (for \mbox{$J_H = 0.1$ eV}). This is insufficient to model even ultrafast optical manipulation of magnetism requiring simulation times \mbox{$\sim 10$ fs}~\cite{Siegrist2019}, and it is much further away  from current-driven magnetization dynamics via spin torque which occurs on $\sim 1$ ns time scales~\cite{Ralph2008,Berkov2008}.

\subsection{Range of validity of truncated HP transformation for  nonequilibrium interacting system of electrons and magnons}\label{sec:validity_em}

In this Section, we repeat the same analysis as in Sec.~\ref{sec:validity_mm}---but with electron--localized-spin or, equivalently electron-magnon---interaction turned on within 1D quantum many-body system composed of $N=3$ sites  [Fig.~\ref{fig:fig1}(b)]. These sites host spin-$\frac{5}{2}$ localized spins interacting with half-filled ($N_e = 3$) tight-binding electrons via the $sd$ exchange interaction~\cite{Cooper1967} of strength $J_\mathrm{sd}$ as encoded by Eq.~\eqref{eq:sd_ham}. 

At $t=0$, we fully flip the localized spin-$\frac{5}{2}$ on site $i=1$ to initiate nonequilibrium dynamics. From the viewpoint of HP transformation, this is equivalent to introducing $N_\mathrm{mag} = 5$ HP bosons on site $i=1$, and thus, the initial quantum many-body state is given by
\begin{equation}\label{eq:fivehpbosons}
    \ket{\Psi(0)} = (\hat{a}_1^\dagger)^5\hat{c}^\dagger_{2\uparrow} \hat{c}^\dagger_{1\uparrow}\hat{c}^\dagger_{1\downarrow}\ket{0},
\end{equation}
in the notation of second-quantization formalism. Here $\ket{0}$ is the vacuum state of electrons and HP bosons combined. Figure~\ref{fig:fig6}(a) with $J_\mathrm{sd}$ = 0 serves as a reference. When electron-magnon interaction is  turned on (\mbox{$J_\mathrm{sd} = 1$ eV}) in  Fig.~\ref{fig:fig6}(b),  $\langle \hat{S}_1^z\rangle (t)$ computed by  truncated HP transformation follows the exact result for longer times \mbox{$t \lesssim 15$ fs} than 
in Fig.~\ref{fig:fig6}(a). This is explained by Fig.~\ref{fig:fig6}(c) which shows that the total number of magnons as a function of time, $N_\mathrm{mag}(t) = \sum_{i=1}^N \bra{\Psi(t)}\hat{n}_i\ket{\Psi(t)}$, is  reduced in the course of quantum time evolution. Therefore, this leads to fewer magnon-magnon scattering events which facilitates accurate tracking over longer time intervals of nonequilibrium dynamics of localized spins by truncated HP transformation, in accord with Fig.~\ref{fig:fig5}. The lost magnons in Fig.~\ref{fig:fig6}(c) are absorbed by the electronic subsystem and mediate transfer of spin angular momentum between the subsystems of electrons and localized spins, while the total $z$-spin remains conserved [Eq.~\eqref{eq:sz_sym}].

Furthermore, Figs.~\ref{fig:fig6}(d) and ~\ref{fig:fig6}(e), as the counterpart of Figs.~\ref{fig:fig6}(a) and ~\ref{fig:fig6}(b), respectively, demonstrate that electron-boson interacting Hamiltonian can track exact time evolution if truncated HP transformation is replaced by resummed HP transformation in  Eq.~\eqref{eq:hp_resum}. That is, when $N_\mathrm{max} = 3$ is used in Eq.~\eqref{eq:hp_resum},  there is disagreement between the two calculations of 
$\langle \hat{S}_1^z \rangle (t)$---compare resummed HP transformation (magenta solid line) vs. the exact one (black solid line)---but increasing $N_\mathrm{max} = 5$ in Eq.~\eqref{eq:hp_resum} ensures that both methods match perfectly. 

Thus, Figs.~\ref{fig:fig6}(d) and ~\ref{fig:fig6}(e) with properly chosen $N_\mathrm{max}$ motivate us to derive electron-boson Hamiltonian 
\begin{equation}\label{eq:exacteb}
	\hat{H} = \hat{H}_e + \hat{H}_\mathrm{mag} + \hat{H}_\mathrm{mag-mag} + \hat{H}_{e-\mathrm{mag}} 
\end{equation}
as the  {\em exact} mapping of the original electron--localized-spin Hamiltonian in Eq.~\eqref{eq:ham_sys}. The former is required for equilibrium or nonequilibrium MBPT~\cite{Stefanucci2013,Schlunzen2020} which can handle~\cite{Mahfouzi2014} systems in two- or three-dimensions composed of large number of sites  $N \gg 1$---that is, the problems where exact diagonalization~\cite{Wang2019} or (time-dependent) density matrix renormalization group~\cite{White2004,Schmitteckert2004,Daley2004,Feiguin2011} (suitable for $N \gg 1$ but only in quasi-1D~\cite{Stoudenmire2012}) are inapplicable. Here the terms in  Eq.~\eqref{eq:exacteb} are given by 

\begin{widetext}
\begin{subequations}\label{eq:em_ham}
\begin{eqnarray}
    \hat{H}_e 
    &=& 
    -\gamma\sum_{\braket{ij}}\hat{\psi}^\dagger_i\hat{\psi}_j, \\    
    \hat{H}_\mathrm{mag} 
    &=& 2SJ_H \sum_{i=1}^N\hat{a}_i^\dagger \hat{a}_i  -J_H S \sum_{\braket{ij}} \big(\hat{a}_i^\dagger \hat{a}_j + \hat{a}_j^\dagger \hat{a}_i \big), \\
    \hat{H}_\mathrm{mag-mag}
    &=& -J_H \sum_{\braket{ij}}(\hat{a}_i^\dagger \hat{a}_i)(\hat{a}_j^\dagger \hat{a}_j)
    \nonumber \\
    &&- J_H S\sum_{\braket{ij}}\sum_{n=0}^{N_\mathrm{max}}\sum_{m=0}^{N_\mathrm{max}}C_nC_m(1 - \delta_{n0}\delta_{m0})\big[
    (\hat{a}_i^\dagger)^n(\hat{a}_i)^n\hat{a}_i\hat{a}_j^\dagger (\hat{a}_j^\dagger)^m (\hat{a}_j)^m + \hat{a}_i^\dagger (\hat{a}_i^\dagger)^n(\hat{a}_i)^n(\hat{a}_j^\dagger)^m(\hat{a}_j)^m\hat{a}_j \big], \\
    \hat{H}_{e-\mathrm{mag}}  &=& \sqrt{2}S J_\mathrm{sd}\sum_{i=1}^N \sum_{n=0}^{N_\mathrm{max}}C_n\big[\hat{\psi}_{i\uparrow}^\dagger \hat{\psi}_{i\downarrow} \hat{a}_i^\dagger(\hat{a}_i^\dagger)^n(\hat{a}_i)^n +  \hat{\psi}_{i\downarrow}^\dagger \hat{\psi}_{i\uparrow} (\hat{a}_i^\dagger)^n(\hat{a}_i)^n \hat{a}_i\big].
\end{eqnarray}
\end{subequations}
\end{widetext}
Their physical meaning is transparent: $\hat{H}_e$ is the tight-binding Hamiltonian of noninteracting electrons; $\hat{H}_\mathrm{mag}$ is the  Hamiltonian of noninteracting HP bosons; $\hat{H}_\mathrm{mag-mag}$ describes various interactions between two (first term in $\hat{H}_\mathrm{mag-mag}$) or more HP bosons; and $\hat{H}_{e-\mathrm{mag}}$ describes electron-boson interactions, such as  absorption or emission of HP bosons  accompanied by electron spin flip as the spin angular momentum is transferred. We note that Eq.~\eqref{eq:em_ham} is much more complex that what is typically used in spintronics literature~\cite{Mahfouzi2014, Tveten2015, Zheng2017, Bender2019, Troncoso2019, Kamra2016}.  Most importantly, it shows that accurate MBPT or diagrammatic Monte Carlo calculations~\cite{Bertrand2019,Bertrand2019a} in the future for interacting electron-magnon system will have to deal with nonlinear~\cite{Marini2018} electron-boson interactions.  

\subsection{Ground state of  interacting system of electrons and magnons}\label{sec:groundstate}
\begin{figure*}
	\centering
	\includegraphics[width=\linewidth]{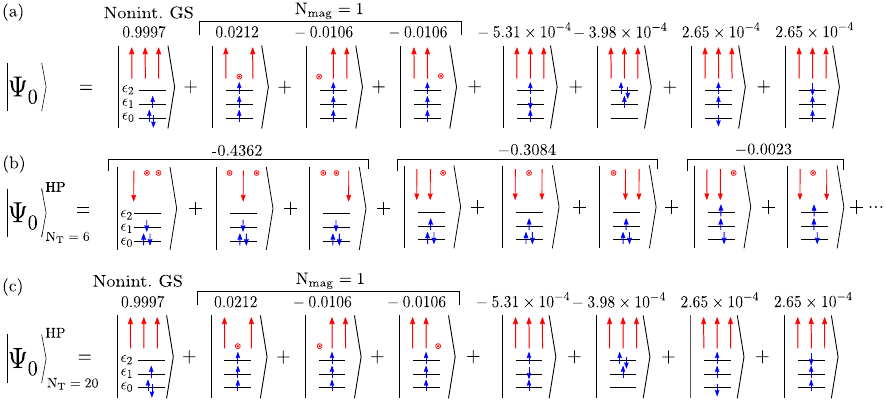}
	\caption{(a) Composition of the {\em exact} GS $\ket{\Psi_0}$ of 1D quantum many-body system, illustrated in Fig.~\ref{fig:fig1}(b) and composed of \mbox{$N=3$} sites hosting spin-1 localized spins, which is computed by exact diagonalization of Hamiltonian in Eq.~\eqref{eq:ham_sys} expressed in terms of localized spin operators. In every ket in the sum, red arrows depict localized spins, whereas blue arrows indicate spin states of $N_e = 3$ electrons distributed among eigenenergy levels \mbox{$\epsilon_0 = -\sqrt{2}$ eV}, \mbox{$\epsilon_1 = 0$ eV}, and   \mbox{$\epsilon_2 = \sqrt{2}$ eV} of  noninteracting single-particle electronic Hamiltonian in Eq.~\eqref{eq:eham_diag}. (b) The composition of {\em approximate} GS evaluated by exact diagonalization of Hamiltonian in Eq.~\eqref{eq:ham_sys} whose localized spins are mapped to bosonic operators via the truncated HP transformation  (Sec.~\ref{sec:hptruncated}) with truncation number $N_T = 6$. Panel (c) shows that in order to match the composition of the exact GS from panel (a) requires to increase $N_T = 20$ in truncated HP transformation. Numbers on the top of each ket are coefficients in their linear superpositions comprising the respective GS. In all panels we set \mbox{$J_H = J_\mathrm{sd} = 0.2$ eV}. Note that the first ket in panels (a) and (c) is noninteracting GS for a system where electrons and localized spins are decoupled by using $J_\mathrm{sd}=0$.}
	\label{fig:fig7}
\end{figure*}
\begin{figure}
	\centering
	\includegraphics[width=\linewidth]{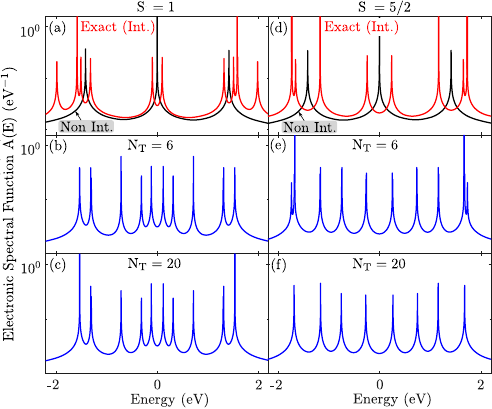}
	\caption{(a) The {\em exact} electronic spectral function $A(E)$ (red lines) in Eq.~\eqref{eq:ele_spec_f} for 1D quantum many-body system,  illustrated in Fig.~\ref{fig:fig1}(b) and  comprised  of $N=3$ sites hosting \mbox{spin-1} localized spins, is evaluated using localized spin operators in Hamiltonian in Eq.~\eqref{eq:ham_sys}. This is compared to {\em approximate} $A(E)$ in panels (b) and (c) evaluated by mapping localized spin operators to bosonic operators via truncated HP transformation  (Sec.~\ref{sec:hptruncated}) with truncation numbers $N_T= 6$ and $N_T=20$, respectively. Panels (d)--(f) show counterpart information to panels (a)--(c), but for spin-$\frac{5}{2}$ localized spins on each of $N=3$ sites. In all panels we set \mbox{$J_H  = J_\mathrm{sd} =0.2$ eV}, except for black curves in panels (a) and (d) for which electron--localized-spin interaction is turned off i.e., $J_\mathrm{sd} =0$.}
	\label{fig:fig8}
\end{figure}

It is also instructive to explore the structure of the exact quantum many-body {\em ground} state (GS) of conduction electrons plus localized spins in {\em equilibrium} (GS implies zero temperature as well), as described by Hamiltonian in Eq.~\eqref{eq:ham_sys} in terms of the original localized spin operators; as well as to find out how many terms of truncated HP transformation (Sec.\ref{sec:hptruncated}) need to be retained in order to obtain the same ground state by exact diagonalization of the Hamiltonian of the same system but expressed in terms of electronic and bosonic operators. In this Section, comparison of ground states in two methods is performed for  the system depicted in Fig.~\ref{fig:fig1}(b) composed of $N=3$ sites hosting spin-$1$ localized spins, while in Secs.~\ref{sec:spectralelectron} and  \ref{sec:spectralmagnon} we also perform comparison of electronic and magnonic spectral functions, respectively, which require additional information about the {\em excited} quantum many-body states. 

The exact GS is obtained in three steps: (\emph{i})  $\hat{H}$ is represented as a matrix in the basis of eigenstates of $\hat{N}_e$ and $\hat{S}^z_\mathrm{tot}$ to render a block-diagonal matrix as shown in Fig.~\ref{fig:fig2}(c); (\emph{ii}) to ensure half-filling for electrons, the matrix block corresponding to $N_e = 3$ electrons is isolated; (\emph{iii}) This matrix block is diagonalized and the eigenstate with the lowest eigenenergy $E_0$ is identified as the GS. Obviously, if in step (\emph{i}) $\hat{H}$ is expressed directly in terms of localized spin operators [Eq.~\eqref{eq:spin_mat}], then step (\emph{iii}) yields the \emph{numerically exact}  GS $\ket{\Psi_0}$. On the other hand, if $\hat{H}$ is expressed using the truncated HP transformation with a truncation number $N_T$ [Eq.~\eqref{eq:split_ham}], then thus obtained GS $\ket{\Psi_0}^\mathrm{HP}_{N_T}$ is not guaranteed to be the same as $\ket{\Psi_0}$. In particular, we are interested to know what value of $N_T$ ensures that $\ket{\Psi_0}^\mathrm{HP}_{N_T} = \ket{\Psi_0}$.


Figure~\ref{fig:fig7}(a) depicts the numerically exact GS $\ket{\Psi_0}$ as a linear combination~\cite{Frederiksen2004} of many-body kets  where red arrows  denote quantum state of localized spins (using the same notation as introduced in Sec.~\ref{sec:magnons} for spin-1 localized spins) while blue arrows denote  \mbox{spin-$\uparrow$} or \mbox{spin-$\downarrow$} electrons filling three single-particle energy levels \mbox{$\epsilon_0 = -\sqrt{2}$ eV}, $\epsilon_1 = 0$, and \mbox{$\epsilon_2 = \sqrt{2}$ eV} of noninteracting tight-binding Hamiltonian $\hat{H}_e$ [Eq.~\eqref{eq:elec_ham}]. In contrast, we find in Fig.~\ref{fig:fig7}(b) that GS $\ket{\Psi_0}^\mathrm{HP}_{N_T=6}$ evaluated using truncated HP transformation with $N_T = 6$ is entirely different from $\ket{\Psi_0}$ shown in Fig.~\ref{fig:fig7}(a).  Only when the truncation number is increased to $N_T = 20$ in Fig.~\ref{fig:fig7}(c) we find $\ket{\Psi_0}^\mathrm{HP}_{N_T=20} \equiv \ket{\Psi_0}$. 

It is worth examining further the structure of the exact GS $\ket{\Psi_0}$ in Fig.~\ref{fig:fig7}(a). Its many-body eigenenergy is \mbox{$E_0 = -3.33$ eV} while the other quantum numbers [Eq.~\eqref{eq:qnumbers}] are $N_e = 3$ and $S^z_\mathrm{tot} = 3.5$. The largest contribution (greater than $99\%$) to $\ket{\Psi_0}$ comes from the first term on the RHS in Fig.~\ref{fig:fig7}(a) where $N_e = 3$ electrons fill up the single-particle energy levels  $\epsilon_0$, $\epsilon_1$, and $\epsilon_2$ of noninteracting Hamiltonian $\hat{H}_e$ [Eq.~\eqref{eq:elec_ham}] in accord with the Pauli exclusion principle while the localized spins are in the ferromagnetic configuration. In the absence of electron--localized-spin interaction, the first term on the RHS would be the only one. Thus, interactions give rise to three states [indicated by horizontal overline in Fig.~\ref{fig:fig7}(a)] where $N_\mathrm{mag} = 1$ HP boson is created on one of the three sites. This HP boson is actually emitted when the spin-$\downarrow$ electron in eigenenergy level $\epsilon_0$  undergoes a spin-flip process and emerges as a spin-$\uparrow$ electron in eigenenergy level $\epsilon_2$. This process respects conservation of total \mbox{$z$-spin} encoded by Eq.~\eqref{eq:sz_sym}. The remaining four kets on the RHS of Fig.~\ref{fig:fig7}(a) are purely electronic excitations where electrons  are excited 
among eigenenergy levels $\epsilon_0$, $\epsilon_1$, and $\epsilon_2$ but are not accompanied by any spin-flip process of localized spins.

\subsection{Electronic spectral function in interacting system of electrons and magnons} \label{sec:spectralelectron}
For the same system considered in Sec.~\ref{sec:groundstate}, Fig.~\ref{fig:fig8}(a)--(c) compares the electronic spectral function $A(E)$ [Eq.~\eqref{eq:ele_spec_f}] evaluated from truncated HP transformation vs. the exact one evaluated using localized spin operators. To set a reference point, 
in Fig.~\ref{fig:fig8}(a) we first consider the noninteracting electronic spectral function (black line) when electron--localized-spin interaction is 
turned off ($J_\mathrm{sd} = 0$). For such a case, the available single-particle states are simply the eigenstates of the noninteracting tight-binding  Hamiltonian  $\hat{H}_e$ [Eq.~\eqref{eq:elec_ham}] with single-particle energy levels $\epsilon_0$, $\epsilon_1$, and $\epsilon_2$, so that  $A(E)$  consists of sharp peaks centered at $\epsilon_0$, $\epsilon_1$, and $\epsilon_2$. Upon turning on electron--localized-spin interaction   (\mbox{$J_\mathrm{sd} = 0.2$ eV}), the exact $A(E)$  (red line) evaluated using localized spin operators  is modified to exhibit peak splitting [with respect to black line reference result when  $J_\mathrm{sd} = 0$] at energies $\epsilon_0$, $\epsilon_1$, and $\epsilon_2$. Also, few additional peaks around single-particle energy levels $\epsilon_0$ and $\epsilon_2$ emerge.

In Fig.~\ref{fig:fig8}(b), we compute $A(E)$ using truncated HP transformation with a truncation number $N_T=6$. Although it reproduces the peak-splitting near $\epsilon_0$,  $\epsilon_1$ and  $\epsilon_2$, it exhibits several additional peaks that are absent in the exact result in Fig.~\ref{fig:fig8}(a). This discrepancy can be understood as follows. The function $A(E)$ depends on the exact GS $\ket{\Psi_0}$ through Eqs.~\eqref{eq:ele_spec_f} and ~\eqref{eq:ele_weight}. However, for  $N_T=6$ Fig.~\ref{fig:fig7}(b) demonstrates $\ket{\Psi_0}^\mathrm{HP}_{N_T = 6} \neq \ket{\Psi_0}$. At first sight, it appears that the same argument should produce exact $A(E)$ in Fig.~\ref{fig:fig8}(c) using $N_T=20$ because 
$\ket{\Psi_0}^\mathrm{HP}_{N_T = 20} \equiv \ket{\Psi_0}$ in Fig.~\ref{fig:fig7}(c). However, the discrepancy between exact $A(E)$ (red line) in Fig.~\ref{fig:fig8}(a) and blue line in Fig.~\ref{fig:fig8}(c) is explained by Eqs.~\eqref{eq:ele_spec_f} and ~\eqref{eq:ele_weight} where $A(E)$ depends  both on GS $\ket{\Psi_0}$ and excited many-body states $\ket{\Psi_k}$ [Eq.~\eqref{eq:ele_weight}] for which truncation number $N_T = 20$ appears to be {\em insufficient}. The repeated analysis from Fig.~\ref{fig:fig8}(a)--(c) for spin-$1$ localized spins, but by using spin-$\frac{5}{2}$ localized spins in Fig.~\ref{fig:fig8}(d)--(f), shows that requirement of large $N_T = 20$ cannot be bypassed by increasing the value of localized spins which make them more ``classical-like''~\cite{Wieser2015,Stahl2017,Gauyacq2014}.

\subsection{Magnonic spectral function in interacting system of electrons and magnons} \label{sec:spectralmagnon}
\begin{figure*}
	\centering
	\includegraphics[width=\linewidth]{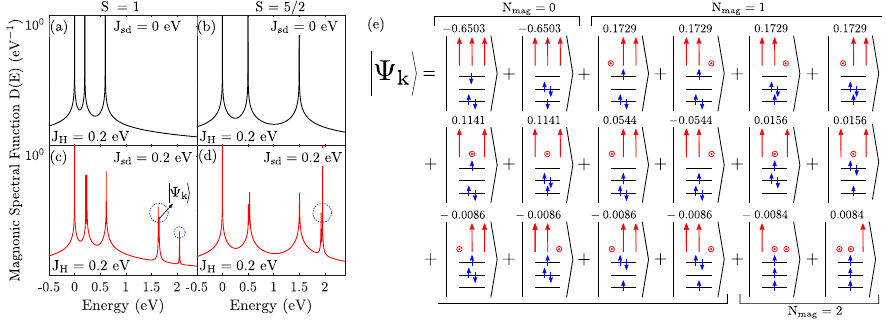}
	\caption{(a)--(d) The {\em exact} magnonic spectral function $D(E)$ in Eq.~\eqref{eq:mag_spec_f} evaluated by mapping localized spin operators to HP bosons in a numerically exact fashion via Eq.~\eqref{eq:hp} for 1D quantum many-body system  illustrated in Fig.~\ref{fig:fig1}(b) comprised  of $N=3$ sites 
	hosting (a),(c) spin-$1$ or (b),(d) spin-$\frac{5}{2}$ localized spins. The electron--localized-spin interaction is turned off  (\mbox{$J_\mathrm{sd} = 0$ eV}) in 
	panels (a) and (b), or turned on (\mbox{$J_\mathrm{sd} = 0.2$ eV}) in panels (c) and (d), while keeping  \mbox{$J_H =  0.2$ eV} in all panels. In panels (c) and (d), dotted circle mark additional peaks in $D(E)$ originating from excited states $|\Psi_k\rangle$, as encoded by Eq.~\eqref{eq:mag_spec_f}. (e) Composition of specific excited quantum many-body state $|\Psi_k\rangle$ for spin-$1$ case which is responsible for peak in $D(E)$ near \mbox{$E=1.6$ eV} in panel (c)---this state is a superposition of kets with zero magnons ($N_\mathrm{mag}=0$) and solely electronic excitations; followed by kets with one magnon excitation ($N_\mathrm{mag}=1$); and then states with two magnons excitations  ($N_\mathrm{mag}=2$). Numbers on the top of each ket are coefficients in their linear superposition leading to $|\Psi_k\rangle$.}
	\label{fig:fig9}
\end{figure*}

The {\em exact} HP transformation in Eq.~\eqref{eq:hp} makes it possible to define magnonic spectral function $D(E)$ in Eq.~\eqref{eq:mag_spec_f} and compute it without any approximations  by numerically evaluating  square root of matrices in Eq.~\eqref{eq:hp}. Using the same systems of spin-$1$ or spin-$\frac{5}{2}$ localized spins that are studied in Fig.~\ref{fig:fig8}, we first establish a reference magnonic spectral function by computing  $D(E)$ in Figs.~\ref{fig:fig9}(a) and \ref{fig:fig9}(b) with electron--localized-spin interaction turned off ($J_\mathrm{sd} = 0$). Such reference $D(E)$ [Fig.~\ref{fig:fig9}(a)] exhibits three peaks at energies \mbox{$E=0$ eV}, \mbox{$E = 0.2$ eV}, and \mbox{$E=0.6$ eV} which correspond to available states in the presence of solely localized-spin--localized-spin (or equivalently magnon-magnon) interactions. Conversely, when we turn on \mbox{$J_\mathrm{sd} = 0.2$ eV} in Figs.~\ref{fig:fig9}(c) and ~\ref{fig:fig9}(d), we find: (\emph{i}) the original noninteracting peaks remain largely intact, except for the one near \mbox{$E=0.2$ eV} which undergoes a tiny splitting; (\emph{ii}) far away the original peaks, $D(E)$ exhibits new additional peaks (marked by dotted circles) near energies \mbox{$E=1.6$ eV} and \mbox{$E=2.1$ eV}. Analogous features are observed for spin-$\frac{5}{2}$ localized spins when switching from \mbox{$J_\mathrm{sd} = 0$} in Fig.~\ref{fig:fig9}(b) to \mbox{$J_\mathrm{sd} \neq 0$} in Fig.~\ref{fig:fig9}(d). 

We note that similar additional peaks in magnonic spectral function, generated by turning on electron-magnon interaction,  were previously observed in MBPT calculations~\cite{Mahfouzi2014} despite being based on resummation of an infinite class of selected diagrams---in contrast, calculations in Figs.~\ref{fig:fig9}(c) and ~\ref{fig:fig9}(d) are nonperturbative and, therefore, correspond to all diagrams being summed to infinite order. These additional peaks in $D(E)$ computed by MBPT were interpreted in Feynmann diagrammatic language as quasibound states of magnons dressed by the cloud of electron-hole pair excitations. Also, MBPT calculations of Refs.~\cite{Mahfouzi2014,Woolsey1970} find much smaller modification of electronic $A(E)$ upon tuning on  electron-magnon interaction. This is explained by magnons being in the strongly interacting regime vs. electrons being in the weakly interacting regime due to~\cite{Mahfouzi2014} $J_\mathrm{sd}$ divided by the bandwidth of noninteracting magnons being much larger than $J_\mathrm{sd}$ divided by the bandwidth of noninteracting electrons. 

To clarify the origin of these peaks further in the context of our exact nonperturbative calculations in Figs.~\ref{fig:fig9}(c) and ~\ref{fig:fig9}(d), we focus on the peak near \mbox{$E=1.6$ eV} in Fig.~\ref{fig:fig9}(c). This peak is due to many-body excited state $\ket{\Psi_k}$ whose composition is given explicitly in Fig.~\ref{fig:fig9}(e). This state has a nonzero ``weight''  $Q_k^+ = 0.006$ in Eq.~\eqref{eq:mag_weight}. Although, the value of $Q_k^+$ appears to be small, it contributes about $2\%$ in the sum rule in Eq.~\eqref{eq:mag_sum_rule} and thus it cannot be ignored.  Interestingly, Fig.~\ref{fig:fig9}(e) reveals that this specific $\ket{\Psi_k}$ is a linear superposition of states with $N_\mathrm{mag} = 0$, $1$, or $2$ HP bosons.

\subsection{Entanglement entropy of ground and excited states of interacting  system of electrons and magnons}\label{sec:entanglement}

All three different version of the GS $\ket{\Psi_0}$ in Fig.~\ref{fig:fig7}, as well as selected excited state $\ket{\Psi_k}$ shown in Fig.~\ref{fig:fig9}(e), are examples of pure but {\em entangled} quantum many-body states~\cite{Chiara2018}. In particular, these states encodes entanglement between electronic and localized-spins subsystems. The von Neumann entanglement entropy~\cite{Chiara2018} for electronic or localized-spins subsystems of the total bipartite system are identical, $\mathcal{S}_e=\mathcal{S}_\mathrm{lspins}$, and can be computed from the reduced density matrix $\hat{\rho}_e$
\begin{equation}\label{eq:entropy}
	\mathcal{S}_e  = -\mathrm{Tr}[\hat{\rho}_e \ln \hat{\rho}_e],
\end{equation} 
where the (improper) mixed quantum state of the electronic subsystem is described by reduced density matrix
\begin{equation}\label{eq:entropy}
	\hat{\rho}_e  = \mathrm{Tr}_\mathrm{lspins} |\Psi\rangle \langle \Psi|,
\end{equation}
which is obtained by partial trace of the pure state density matrix, $|\Psi\rangle \langle \Psi|$,  over the basis of states in $\mathcal{H}_\mathrm{lspins}$. For example, $\mathcal{S}_e^0=5.6 \times 10^{-3}$ for the exact GS in Fig.~\ref{fig:fig7}(a), which means that this many-body entangled state is quite close to separable (characterized by $\mathcal{S}_e \equiv 0$) noninteracting (i.e., for $J_\mathrm{sd}=0$) GS as the first term depicted in Fig.~\ref{fig:fig7}(a). On the other hand, $\mathcal{S}_e^0=0.604$ for the GS in Fig.~\ref{fig:fig7}(b) which is incorrect [unlike the correct GS in Fig.~\ref{fig:fig7}(c) which matches the exact GS in Fig.~\ref{fig:fig7}(a)] due to too small $N_T$ employed in truncated HP transformation [Sec.~\ref{sec:hptruncated}]. Note that selected excited many-body entangled state $|\Psi_k\rangle$ analyzed in Fig.~\ref{fig:fig9}(e) has much larger  $\mathcal{S}_e^k=0.467$.

\subsection{Diagonal and off-diagonal elements of time-dependent electronic and magnonic lesser Green functions}\label{sec:diag_offdiag}
\begin{figure*}
	\centering
	\includegraphics[width=\linewidth]{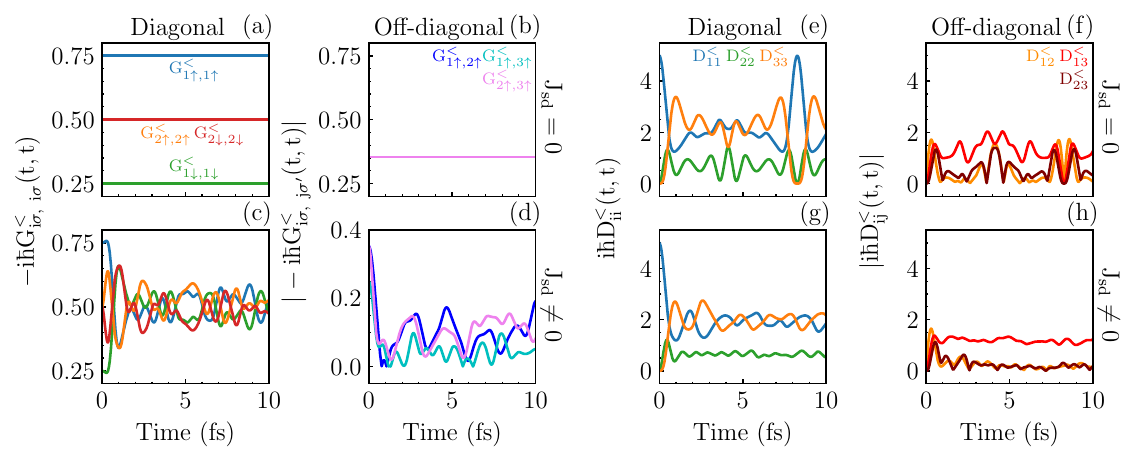}
	\caption{Time evolution of (a) diagonal and (b) magnitude of complex off-diagonal elements of the electronic lesser GF $G^<_{i\sigma, j \sigma'}(t, t)$ [Eq.~\eqref{eq:ele_GF_less}] in the site basis for the system depicted in Fig.~\ref{fig:fig1}(b) composed of $N=3$ sites hosting spin-$\frac{5}{2}$ localized spins and with electron-magnon interaction turned-off ($J_\mathrm{sd} = 0$). Panels (c) and (d) are their counterparts when the electron-magnon interaction is turned on (\mbox{$J_\mathrm{sd} = 1$ eV}). Panels (e)--(h) show the same information as panels (a)--(d), but for magnonic lesser GF $D^<_{ij}(t, t)$ [Eq.~\eqref{eq:mag_GF_less}] in the site basis. Note that some of the off-diagonal elements are not explicitly shown because they are either zero or identical to the ones plotted in panels (b),(d) and panels (f),(h).}
	\label{fig:fig10}
\end{figure*}

The time-dependent electronic lesser GF $G^<_{i\sigma, j\sigma'}(t, t')$ in Eq.~\eqref{eq:ele_GF_less} generally depends~\cite{Stefanucci2013,Schlunzen2020} on two time arguments, $t$ and $t'$. At equal times $t'=t$, it yields electronic one-particle {\em nonequilibrium} density matrix~\cite{Stefanucci2013,Petrovic2018,Gaury2014,Bajpai2019b}
\begin{equation}
    \bm{\rho}(t) = -i\hbar \bold{G}^<(t, t')\big|_{t'=t}.
\end{equation}
Its diagonal elements in, e.g., coordinate (or site for discrete lattice) representation contain information about the time-dependent electronic charge and spin density~\cite{Petrovic2018}, whereas the off-diagonal elements encode quantum-mechanical interference effects~\cite{Bajpai2019b} and measure the degree of quantum coherence~\cite{Schlosshauer2005}. To illustrate their time evolution, we use the same 1D quantum many-body system employed in Fig.~\ref{fig:fig6} where localized spin-$\frac{5}{2}$ on site  $i=1$ is completely flipped [i.e., $N_\mathrm{mag} = 5$ HP bosons are introduced on site $i=1$ via Eq.~\eqref{eq:fivehpbosons}] to initiate nonequilibrium dynamics. 

Figure~\ref{fig:fig10}(a)--(d) shows the ensuing time evolution for the diagonal elements, $-i\hbar G^<_{i\sigma, i \sigma}(t, t)$, as well as for the off-diagonal elements,  $-i\hbar G^<_{i\sigma, j \sigma'}(t, t)$. In order to establish a reference result, we turn  electron--localized-spin interaction off ($J_\mathrm{sd} = 0$) in Figs.~\ref{fig:fig10}(a) and ~\ref{fig:fig10}(b), which trivially leads to all elements being time-independent because for $J_\mathrm{sd} = 0$ the quantum state of the electronic subsystem is an eigenstate of the electronic Hamiltonian $\hat{H}_e$ [Eq.~\eqref{eq:elec_ham}]. 

Conversely, Figs.~\ref{fig:fig10}(c) and ~\ref{fig:fig10}(d) use \mbox{$J_\mathrm{sd} = 1$ eV} which leads to nontrivial time dependence of both diagonal and off-diagonal elements of $-i\hbar\bold{G}^<(t, t)$. Interestingly, the diagonal elements, $-i\hbar G^<_{i\sigma, i\sigma}(t, t)$ in Fig.~\ref{fig:fig10}(c)  satisfy $-i\hbar[G^<_{i\uparrow, i\uparrow}(t, t) + G^<_{i\downarrow, i\downarrow}(t, t)] = Q_i$ with $Q_i$ being the total electronic density on site $i$, are time-independent. This means that no charge currents flows between sites $i$ and $j$. Instead, population of electrons with spin \mbox{$\sigma=\uparrow,\downarrow$} on site $i$ exchanges solely between spin \mbox{$\sigma=\uparrow, \downarrow$} states at that site. This is also accompanied by time evolution of the off-diagonal elements $-i\hbar G^<_{i\sigma, j\sigma'}(t, t)$ in Fig.~\ref{fig:fig10}(d). 

The off-diagonal elements of the lesser GF are also required to calculate many-body lesser self-energy  $\bm{\Sigma}^<(t_1, t_2)$~\cite{Stefanucci2013,Schlunzen2020}, which is connected to lesser GF in a self-consistent fashion via the Keldysh equation
\begin{equation}\label{eq:gless_sc}
    \bold{G}^<(t, t') =
    \int\displaylimits_{-\infty}^{+\infty}
    \int\displaylimits_{-\infty}^{+\infty}
    dt_1 dt_2
    \bold{G}^r(t, t_1)\bold{\Sigma}^<(t_1, t_2)\bold{G}^a(t_2, t').
\end{equation}
Equation~\eqref{eq:gless_sc} encapsulates time evolution of quantum many-body systems in terms of solely one-particle quantities. Here $\bold{G}^a(t,t')=[\bold{G}^r(t',t)]^\dagger$ is the advanced GF. A self-consistent solution to Eq.~\eqref{eq:gless_sc} can yield exact many-body lesser self-energy. Alternatively, one can systematically approximate it~\cite{Mahfouzi2014}  using the so-called ``conserving approximations''~\cite{Mera2016} in MBPT. One such ``conserving approximation'' for the lesser self-energy of electron-boson interacting systems is the so-called self-consistent Born approximation (SCBA)~\cite{Mahfouzi2014,Frederiksen2007,Lee2009,Mera2016}. The SCBA ensures charge 
conservation in nonequilibrium~\cite{Frederiksen2004}, and in steady-state nonequilibrium one can Fourier transform ${\Sigma}_{ij}^<(t_1-t_2)$ to 
energy domain and operate with $\Sigma_{ij}^<(E)$. 
    
To reduce computational complexity~\cite{Frederiksen2007} of calculations of $\Sigma^<_{ij}(E)$ and enable simulations of devices containing large number of atoms, the ``local self-energy'' approximation is often employed~\cite{Luiser2009,Rhyner2014,Cavassilas2016,Bescond2018} when modeling inelastic scattering of electrons and bosons. In this approximation, one assumes $|\Sigma^<_{ii}(E)| \gg |\Sigma^<_{ij}(E)|$, i.e., the off-diagonal elements of self-energy are minuscule when compared to the diagonal ones, and thus, one can set them to zero. This is done in conjunction with discarding the off-diagonal elements of the electronic lesser GF i.e., $G^<_{ij}(E) \approx G^<_{ii}(E)\delta_{ij}$. 

Using our numerically exact electronic lesser GF in Fig.~\ref{fig:fig10}(c),(d), we can explicitly check if the ``local self-energy'' approximation~\cite{Luiser2009,Rhyner2014,Cavassilas2016,Bescond2018} is warranted for electron-magnon realization of electron-boson quantum many-body system. The off-diagonal elements of the lesser GF in Fig.~\ref{fig:fig10}(d) are not minuscule, but are  instead approximately one-fifth of the diagonal elements in Fig.~\ref{fig:fig10}(c). Therefore, ``local self-energy'' approximation cannot be justified in the case of many-body electron-magnon interacting systems. Figure~\ref{fig:fig10}(e)--(h) shows the counterpart of Fig.~\ref{fig:fig10}(a)--(d) but for the lesser GF of HP bosons. Here  the off-diagonal elements of the bosonic lesser GF, $\mathbf{D}^<(t,t)$ in Eq.~\eqref{eq:mag_GF_less}, at equal times (i.e., of bosonic one-particle nonequilibrium density matrix) are always comparable to the diagonal ones independently of whether the electron--localized-spin interaction $J_\mathrm{sd}$ is turned off [Fig.~\ref{fig:fig10}(e),(f)] or turned on [Fig.~\ref{fig:fig10}(g),(h)].

\section{Conclusions}\label{sec:conclusions}

By applying numerically exact diagonalization techniques to two versions of the Hamiltonian of quantum many-body system  of conduction electrons interacting with localized spins that are widely used in spintronics and magnonics, we compare predictions from these two Hamiltonians for: ground state and spectral functions extracted from the retarded GF in {\em equilibrium}; and time evolution of the expectation values of localized spin operators  and lesser GF in {\em nonequilibrium}. The two Hamiltonians, describing  systems illustrated in Fig.~\ref{fig:fig1} chosen as 1D and small in order to make  calculations  tractable, differ in their treatment of localized quantum spins---they are described by either finite-size matrices of the original spin operators or infinite matrices of bosonic operators after the original localized spin operators are mapped to bosonic ones using the popular HP transformation. The truncation [Sec.~\ref{sec:hptruncated}] of HP transformation is always done to make diagrammatic MBPT~\cite{Stefanucci2013,Mahfouzi2014} or Monte Carlo~\cite{Bertrand2019,Bertrand2019a} calculations possible, but mapping of finite size to infinite matrices necessarily requires some approximations which can lead to spurious effects in equilibrium (Fig.~\ref{fig:fig8})  or incorrect time evolution  (Figs.~\ref{fig:fig4}--\ref{fig:fig6}) out of equilibrium. Our conclusions are summarized as follows:

\begin{enumerate}
	
\item  For quantum many-body systems composed of localized spins alone, Fig.~\ref{fig:fig4} shows that as more interacting HP bosons are introduced into the system, progressively larger number of terms $N_T$  is required in truncated HP transformation to incorporate multi-magnon interactions and accurately track the nonequilibrium dynamics of localized spins. Figure~\ref{fig:fig5} shows that the breakdown-time $t_\mathrm{break}$ for truncated HP transformation follows $t_\mathrm{break} \propto \exp(pN_T)$. Although, the exponential dependence of $t_\mathrm{break}$ on the truncation number $N_T$ is favorable, the reasonable value of $N_T = 1$--$5$ typically used in practical calculations does not allow one to track dynamics beyond $\sim 15$ fs time scale which is insufficient for ultrafast~\cite{Siegrist2019} or spin torque applications~\cite{Ralph2008,Berkov2008}.

\item When electrons are introduced and  electron--localized-spin interaction is turned on, Fig.~\ref{fig:fig6}(a)--(c) shows that $N_T$ required to accurately track nonequilibrium dynamics of localized spins is actually reduced due to the transfer of spin angular momentum between the two subsystems, which  effectively reduces the total number of interacting magnons within the localized spin subsystem. Furthermore, Figs.~\ref{fig:fig6}(d) and \ref{fig:fig6}(e) show that the recently introduced~\cite{Vogl2020} resummed HP transformation [Sec.~\ref{sec:hpresum}] makes it possible to {\em completely evade artifacts} of the usual truncated HP transformation. However, the electron-magnon Hamiltonian furnished by it in Eq.~\eqref{eq:em_ham} is much more complex for MBPT and diagrammatic Monte Carlo calculations than previously used electron-magnon Hamiltonians~\cite{Mahfouzi2014} based on low-order truncated HP transformation.

\item Figure~\ref{fig:fig7} reveals how  truncated HP transformation with a small truncation number  [such as $N_T = 6$ in Fig.~\ref{fig:fig7}(b)] produces an incorrect GS of the interacting electron-magnon system. Only when truncation number is increased  [such as to $N_T = 20$ in Fig.~\ref{fig:fig7}(c)], exact diagonalization of electron-boson Hamiltonian reproduces the exact GS obtained by diagonalizing the original electron--localized-spin-operators Hamiltonian  [Fig.~\ref{fig:fig7}(a)]. However, even large truncation number [such as $N_T = 20$ in Fig.~\ref{fig:fig8}(c),(f)] does not ensure that correct electronic  spectral function can be obtained from electron-boson Hamiltonian due to the fact that spectral functions depends [Eq.~\eqref{eq:ele_spec_f}] on {\em both} the GS and excited quantum many-body states. 

\item The magnonic spectral function can be substantially modified [Fig.~\ref{fig:fig9}(c),(d)] upon introduction of conduction electrons and their interaction with localized spins, even when such interaction appears small for electrons, due to much smaller bandwidth of magnons. That is,  magnons are effectively  pushed into strongly interacting regime, and the new peaks in their spectral function (or ``interacting density of states''~\cite{Balzer2011,Nocera2018}) can be directly related to specific excited quantum many-body states. The structure of excited states [Fig.~\ref{fig:fig9}(a)] reveals superpositions of many-body states in which holes in electronic single particle levels are formed and accompanied by flips of localized spins or, equivalently, creation of one or more virtual HP bosons. 

\item The time evolution of the matrix elements of the lesser GF (electronic or magnonic) at equal times in real-space representation, which yields the one-particle nonequilibrium density matrix in real-space representation, shows that the magnitude of the off-diagonal elements is always comparable to the 
magnitude of the diagonal ones (Fig.~\ref{fig:fig10}). Thus, ``local self-energy'' approximation neglecting the off-diagonal elements, as often employed~\cite{Luiser2009,Rhyner2014,Cavassilas2016,Bescond2018} to enable MBPT modeling of electron-boson systems with large number of atoms, is not  warranted.   

\end{enumerate}
\begin{acknowledgments}
This research was primarily supported by the US National Science Foundation (NSF) through the University of Delaware Materials Research Science and Engineering Center DMR-2011824. The paper has originated from ``Research Projects Based Learning'' implemented within a graduate course {\em PHYS814: Advanced Quantum Mechanics}~\cite{phys814}  at the University of Delaware. 
\end{acknowledgments}



\begin{thebibliography}{10}

\bibitem{Bloch1930}
F. Bloch, Zur Theorie des Ferromagnetismus, Z. Phys. {\bf 61},  206  (1930).

\bibitem{Chumak2015}
A.~V. Chumak, V.~I. Vasyuchka, A.~A. Serga, and B. Hillebrands, Magnon spintronics, Nat. Phys. {\bf 11},  453  (2015).

\bibitem{Wieser2015}
R.~Wieser, Description of a dissipative quantum spin dynamics with a Landau-Lifshitz-Gilbert like damping and complete derivation of the classical
Landau-Lifshitz equation, Euro. Phys. J. B {\bf 88}, 77 (2015). 

\bibitem{Kim2010}
S.-K. Kim, Micromagnetic computer simulations of spin waves in nanometre-scale patterned magnetic elements, J. Phys. D: Appl. Phys. {\bf  43}, 264004 (2010). 

\bibitem{Evans2014}
R.~F.~L. Evans, W.~J. Fan, P.~Chureemart, T.~A. Ostler, M.~O.~A. Ellis, and R.~W. Chantrell, Atomistic spin model simulations of magnetic
nanomaterials, J. Phys.: Condens. Matter {\bf 26}, 103202 (2014).

\bibitem{Zhitomirsky2013}
M. E. Zhitomirsky and A. L. Chernyshev, Colloquium: Spontaneous magnon decays, Rev. Mod. Phys. {\bf 83}, 219 (2013).

\bibitem{Holstein1940}
T. Holstein and H. Primakoff, Field dependence of the intrinsic domain magnetization of a ferromagnet, Phys. Rev. {\bf 58},  1098  (1940).

\bibitem{Mahan2011}
G. D. Mahan, {\em  Condensed Matter in a Nutshell} (Princeton University Press, Princeton, 2011).

\bibitem{Chudnovsky2006} 
E. M. Chudnovsky and J.~Tejada, {\em Lectures on Magnetism} (Rinton Press, Paramus, 2006).

\bibitem{Kim2016}
S.-K. Kim, H.~Ochoa, R. Zarzuela, and Y. Tserkovnyak, Realization of the Haldane-Kane-Mele model in a system of localized spins, Phys. Rev. Lett. {\bf 117}, 227201 (2016). 

\bibitem{Mook2021}
A. Mook, K. Plekhanov, J. Klinovaja, and D. Loss, Interaction-stabilized topological magnon insulator in ferromagnets, Phys.  Rev. X {\bf 11}, 021061 (2021).

\bibitem{Elyasi2020}
M. Elyasi, Y. M. Blanter, and G. E. W. Bauer, Resources of nonlinear cavity magnonics for quantum information, Phys. Rev. B {\bf 101}, 054402 (2020).

\bibitem{Tupitsyn2008}
I. S. Tupitsyn, P. C. E. Stamp, and A. L. Burin, Stability of Bose-Einstein condensates of hot magnons in Yttrium Iron Garnet films, Phys. Rev. Lett. {\bf 100}, 257202 (2008).

\bibitem{Yuan2020}
H. Y. Yuan  and R. A. Duine, Magnon antibunching in a nanomagnet, Phys. Rev. B {\bf 102}, 100402(R) (2020).

\bibitem{Takei2019}
S. Takei, Spin transport in an electrically driven magnon gas near Bose-Einstein condensation: Hartree-Fock-Keldysh theory, Phys. Rev. B {\bf 100}, 134440 (2019).

\bibitem{Radosevic2015}
S. M. Rado\v{s}evi\'{c}, Magnon-magnon interactions in $O(3)$ ferromagnets and equations of motion for spin operators, Ann. Phys. {\bf 362}, 336 (2015).

\bibitem{Vogl2020}
M. Vogl, P. Laurell, H. Zhang, S. Okamoto, and G. A. Fiete, Resummation of the Holstein-Primakoff expansion and differential equation approach to operator square roots, Phys. Rev. Research {\bf 2}, 043243 (2020).

\bibitem{Konig2021}
J. K\"{o}nig and A. Hucht, Newton series expansion of bosonic operator functions, SciPost Phys. {\bf 10}, 007 (2021).

\bibitem{Dyson1956}
F.~J. Dyson, General theory of spin-wave interactions, Phys. Rev. {\bf 102},  1217  (1956).

\bibitem{Hofmann2011}
C. P. Hofmann, Spontaneous magnetization of an ideal ferromagnet: Beyond Dyson's analysis, Phys. Rev. B {\bf 84}, 064414 (2011).

\bibitem{Schuckert2018}
A. Schuckert, A. Pi\~{n}eiro Orioli, and J. Berges, Nonequilibrium quantum spin dynamics from two-particle irreducible functional integral techniques in the Schwinger boson representation, Phys. Rev. B {\bf 98}, 224304 (2018).

\bibitem{Stefanucci2013}
G.~Stefanucci and R.~van Leeuwen, \emph{Nonequilibrium Many-Body Theory of Quantum Systems: A Modern Introduction} (Cambridge University Press, Cambridge, 2013).

\bibitem{Schlunzen2020}
N. Schl\"{u}nzen, S. Hermanns, M. Scharnke, and M. Bonitz, Ultrafast dynamics of strongly correlated fermions-nonequilibrium Green functions and selfenergy approximations, J. Phys.: Condens. Matter {\bf 32}, 103001 (2020). 

\bibitem{Leeuwen2012}
R. van Leeuwen and G. Stefanucci, Wick theorem for general initial states, Phys. Rev. B {\bf 85},  115119  (2012).

\bibitem{Kubo1953}
R. Kubo, The spin-Wave theory as a variational method and its application to antiferromagnetism, Rev. Mod. Phys. {\bf 25}, 344 (1953).

\bibitem{Harris1971} 
A. B. Harris, D. Kumar, B. I. Halperin, and P. C. Hohenberg, Dynamics of an antiferromagnet at low temperatures: Spin-wave damping and hydrodynamics, Phys. Rev. B {\bf 3}, 961 (1971).

\bibitem{Hamer1992}
C. J. Hamer, Z. Weihong, P. Arndt, Third-order spin-wave theory for the Heisenberg antiferromagnet, Phys. Rev. B {\bf 46}, 6276 (1992).

\bibitem{Maleev1958}
S. V. Maleev, Scattering of slow neutrons in ferromagnets, Sov. Phys. JETP {\bf 6}, 776 (1958).

\bibitem{Jordan1928}
P. Jordan and E. Wigner, \"{U}ber das Paulische \"{a}quivalenzverbot, Z. Phys. {\bf 47}, 631 (1928).

\bibitem{Affleck1998}
I. Affleck and J. B. Marston, Large-$n$ limit of the Heisenberg-Hubbard model: Implications for high-$T_c$ superconductors, Phys. Rev. B {\bf 37}, 3774(R) (1988).

\bibitem{Tsvelik1992}
A. M. Tsvelik, New fermionic description of quantum spin liquid state, Phys. Rev. Lett. {\bf 69}, 2142 (1992).

\bibitem{Coleman2000} 
P. Coleman, C. P\'{e}pin, and A. M. Tsvelik, Supersymmetric spin operators, Phys. Rev. B {\bf 62}, 3852 (2000).

\bibitem{Kiselev2000}
M. N. Kiselev and R. Oppermann, Schwinger-Keldysh semionic approach for quantum spin systems, Phys. Rev. Lett. {\bf 85}, 5631 (2000).

\bibitem{Marcuzzi2016}
M. Marcuzzi, J. Marino, A. Gambassi, and A. Silva, Prethermalization from a low-density Holstein-Primakoff expansion, Phys. Rev. B {\bf 94}, 214304 (2016).

\bibitem{Hirsch2013}
J. G. Hirsch, O. Casta\~{n}os, R. L\'{o}pez-Pe\~{n}a, and E. Nahmad-Achar, Virtues and limitations of the truncated Holstein–Primakoff description of quantum rotors, Phys. Scr. {\bf 87}, 038106 (2013).

\bibitem{Mahfouzi2014}
F. Mahfouzi and B.~K. Nikoli\'{c}, Signatures of electron-magnon interaction in charge and spin currents through magnetic tunnel junctions: A nonequilibrium many-body perturbation theory approach, Phys. Rev. B {\bf 90},  045115  (2014).

\bibitem{Tay2013}
T. Tay and L. J. Sham, Theory of atomistic simulation of spin-transfer torque in nanomagnets, Phys. Rev. B {\bf 87}, 174407 (2013). 

\bibitem{Cheng2019}
Y. Cheng, W. Wang, and S. Zhang, Amplification of spin-transfer torque in magnetic tunnel junctions with an antiferromagnetic barrier, Phys. Rev. B {\bf 99}, 104417 (2019). 

\bibitem{Bender2019} 
S. A. Bender, R. A. Duine, and Y. Tserkovnyak, Quantum spin-transfer torque and magnon-assisted transport in nanostructures, Phys. Rev. B {\bf 99}, 024434 (2019).

\bibitem{Okuma2017}
N. Okuma and K. Nomura, Microscopic derivation of magnon spin current in a topological insulator/ferromagnet heterostructure, Phys. Rev. B {\bf 95}, 115403 (2017).

\bibitem{Tveten2015}
E. G. Tveten, A. Brataas, and Y. Tserkovnyak, Electron-magnon scattering in magnetic heterostructures far out of equilibrium, Phys. Rev. B {\bf 92}, 180412(R) (2015).

\bibitem{Zheng2017}
J.~Zheng, S.~Bender, J.~Armaitis, R.~E. Troncoso, and R.~A. Duine, Green's function formalism for spin transport in metal-insulator-metal heterostructures, Phys. Rev. B {\bf 96}, 174422 (2017).

\bibitem{Troncoso2019}
R. E. Troncoso, A. Brataas, and R. A. Duine, Many-body theory of spin-current driven instabilities in magnetic insulators, Phys. Rev. B {\bf 99}, 104426 (2019).

\bibitem{Adachi2011}
H. Adachi, J.-I. Ohe, S. Takahashi, and S. Maekawa, Linear-response theory of spin Seebeck effect in ferromagnetic insulators, Phys. Rev. B {\bf 83}, 094410 (2011).

\bibitem{Kamra2016}
A. Kamra and W. Belzig, Super-Poissonian shot noise of squeezed-magnon mediated spin transport, Phys. Rev. Lett. {\bf 116}, 146601 (2016).

\bibitem{Parvini2020}
T. S. Parvini, V. A. S. V. Bittencourt, and S. V. Kusminskiy, Antiferromagnetic cavity optomagnonics, Phys. Rev. Research {\bf 2}, 022027(R) (2020).

\bibitem{Bender2019a}
S. A. Bender, A. Kamra, W. Belzig, and R. A. Duine, Spin current cross-correlations as a probe of magnon coherence, Phys. Rev. Lett. {\bf 122}, 187701 (2019).

\bibitem{Ralph2008}
D.~Ralph and M.~Stiles, Spin transfer torques, J. Magn. Magn. Mater. {\bf 320}, 1190 (2008).

\bibitem{Petrovic2021}
M. D. Petrovi\'{c}, P. Mondal, A. Feiguin, P. P.~Plech\'{a}\v{c}, and B. K. Nikoli\'{c}, Spintronics meets density matrix renormalization group: Quantum spin-torque-driven nonclassical magnetization reversal and dynamical buildup of long-range entanglement, Phys. Rev. X {\bf 11}, 021062 (2021).

\bibitem{Berkov2008}
D.~V. Berkov and J. Miltat, Spin-torque driven magnetization dynamics: Micromagnetic modeling, J. Magn. Magn. Mater. {\bf 320},  1238  (2008).

\bibitem{Ellis2017}
M.~O.~A. Ellis, M. Stamenova, and S. Sanvito, Multiscale modeling of current-induced switching in magnetic tunnel junctions using {\em ab initio} spin-transfer torques, Phys. Rev. B {\bf 96},  224410 (2017).

\bibitem{Petrovic2018}
M.~D. Petrovi\'{c}, B.~S. Popescu, U.~Bajpai, P.~Plech\'{a}\v{c}, and B.~K. Nikoli\'{c}, Spin and charge pumping by a steady or pulse-current-driven magnetic domain wall: A self-consistent multiscale time-dependent quantum-classical hybrid approach, Phys. Rev. Applied \textbf{10}, 054038 (2018).

\bibitem{Bajpai2019a}
U.~Bajpai and B.~K. Nikoli\'{c}, Time-retarded damping and magnetic inertia in the Landau-Lifshitz-Gilbert equation self-consistently coupled to electronic	time-dependent nonequilibrium Green functions, Phys. Rev. B \textbf{99}, 134409 (2019).

\bibitem{Suresh2020}
A. Suresh, U. Bajpai, and B. K. Nikoli\'{c}, Magnon-driven chiral charge and spin pumping and electron-magnon scattering from time-dependent quantum transport combined with classical atomistic spin dynamics, Phys. Rev. B {\bf 101}, 214412 (2020). 

\bibitem{Suresh2021}
A. Suresh, U. Bajpai, M. D. Petrovi\'{c}, H. Yang, and B. K. Nikoli\'{c}, Magnon- versus electron-mediated spin-transfer torque exerted by spin current across an antiferromagnetic insulator to switch the magnetization of an adjacent ferromagnetic metal, Phys. Rev. Applied {\bf 15}, 034089 (2021).

\bibitem{Bajpai2020}
U. Bajpai and B. K. Nikoli\'{c}, Spintronics meets nonadiabatic molecular dynamics: Geometric spin torque and damping on dynamical classical magnetic texture due to an electronic open quantum system, Phys. Rev. Lett. {\bf 125}, 187202 (2020). 

\bibitem{Stahl2017}
C. Stahl and M. Potthoff, Anomalous spin precession under a geometrical torque, Phys. Rev. Lett. {\bf 119}, 227203 (2017).

\bibitem{Bostrom2019}
E. V. Bostr\"{o}m and C. Verdozzi, Steering magnetic skyrmions with currents: A nonequilibrium Green's functions approach, Phys. Stat. Solidi B {\bf 256}, 1800590 (2019).

\bibitem{Gauyacq2014}
J. P. Gauyacq and N. Lorente, Classical limit of a quantal nano-magnet in an anisotropic environment, Surf. Sci. {\bf 630}, 325 (2014).

\bibitem{Mondal2019}
P. Mondal, U. Bajpai, M. D. Petrovi\'{c}, P. P.~Plech\'{a}\v{c}, and B. K. Nikoli\'{c}, Quantum spin-transfer torque induced nonclassical magnetization dynamics and electron-magnetization entanglement, Phys. Rev. B {\bf 99}, 094431 (2019).

\bibitem{Petrovic2021a}
M. D. Petrovi\'{c}, P. Mondal, A. E. Feiguin, and B. K. Nikoli\'{c}, Quantum spin torque driven transmutation of an antiferromagnetic Mott insulator, Phys. Rev. Lett. {\bf 126}, 197202 (2021).

\bibitem{Mitrofanov2020}
A. Mitrofanov and S. Urazhdin, Energy and momentum conservation in spin transfer, Phys. Rev. B {\bf 102}, 184402 (2020).

\bibitem{Mitrofanov2021}
A. Mitrofanov and S. Urazhdin, Nonclassical spin transfer effects in an antiferromagnet, Phys. Rev. Lett. {\bf 126}, 037203 (2021).

\bibitem{Zholud2017}
A. Zholud, R. Freeman, R. Cao, A. Srivastava, and S. Urazhdin, Spin transfer due to quantum magnetization fluctuations, Phys. Rev. Lett. {\bf 119}, 257201 (2017).

\bibitem{Wang2019}
Y. Wang, J.~P.~Dehollain, F.~Liu, U.~Mukhopadhyay, M.~S.~Rudner, L.~M.~K. Vandersypen, and E.~Demler, {\em Ab initio} exact diagonalization simulation of the Nagaoka transition in quantum dots,  Phys. Rev. B {\bf 100},  155133  (2019).


\bibitem{Parkinson1985}
J. B. Parkinson, J. C. Bonner, G. M\"{u}ller, M. P. Nightingale, and H. W. J. Bl\"{u}te, Heisenberg spin chains: Quantum-classical crossover and the Haldane conjecture, J. Appl. Phys. {\bf 57}, 3319 (1985).

\bibitem{Cooper1967}
R. L. Cooper and E. A. Uehling, Ferromagnetic resonance and spin diffusion in supermalloy, Phys. Rev. {\bf 164}, 662 (1967).

\bibitem{Tsvelik2017}
A.~M. Tsvelik and O.~M. Yevtushenko, Chiral spin order in Kondo-Heisenberg systems, Phys. Rev. Lett. {\bf 119}, 247203 (2017).

\bibitem{Spinelli2014}
A.~Spinelli, B.~Bryant, F.~Delgado, J.~Fern\'{a}ndez-Rossier, and A.~F. Otte, Imaging of spin waves in atomically designed nanomagnets, Nat. Mater. \textbf{10}, 782 (2014).

\bibitem{Loth2012} 
S. Loth, S. Baumann, C. P. Lutz, D. M. Eigler, and A. J. Heinrich, Bistability in atomic-scale antiferromagnets, Science  {\bf 335}, 196 (2012). 

\bibitem{Schumann2010}
R. Schumann and D. Zwicker, The Hubbard model extended by nearest-neighbor Coulomb and exchange interaction on a cubic cluster - rigorous and exact results, Ann. Phys. (Berlin) {\bf 522}, 419 (2010).

\bibitem{Carrascal2015}
D.~J. Carrascal, J. Ferrer, J.~C. Smith, and K. Burke, The Hubbard dimer: a density functional case study of a many-body problem, J. Phys. Condens. Matter {\bf 27},  393001  (2015).

\bibitem{Hermanns2014}
S. Hermanns, N. Schl\"{u}nzen, and M. Bonitz, Hubbard nanoclusters far from equilibrium, Phys. Rev. B {\bf 90}, 125111 (2014).

\bibitem{Sakkinen2015}
N. S\"{a}kkinen, Y. Peng, H. Appel, and R. van Leeuwen, Many-body Green's function theory for electron-phonon interactions: Ground state properties of the Holstein dimer, J. Chem. Phys. {\bf 143}, 234101 (2015).

\bibitem{Sakkinen2015a}
N. S\"{a}kkinen, Y. Peng, H. Appel, and Robert van Leeuwen, Many-body Green's function theory for electron-phonon interactions:
The Kadanoff-Baym approach to spectral properties of the Holstein dimer, J. Chem. Phys. {\bf 143}, 234102 (2015).

\bibitem{Dimitrov2017}
T. Dimitrov, J. Flick, M. Ruggenthaler and A. Rubio, Exact functionals for correlated electron-photon systems, New J. Phys. {\bf 19},  113036 (2017).

\bibitem{Esterlis2018}
I. Esterlis, B. Nosarzewski, E. W. Huang, B. Moritz, T. P. Devereaux, D. J. Scalapino, and S. A. Kivelson, Breakdown of the Migdal-Eliashberg theory: A determinant quantum Monte Carlo study, Phys. Rev. B {\bf 97}, 140501(R) (2018).

\bibitem{Gukelberger2015}
J. Gukelberger, L. Huang, and P. Werner, On the dangers of partial diagrammatic summations: Benchmarks for the two-dimensional Hubbard model in the weak-coupling regime,  Phys. Rev. B {\bf 91}, 235114 (2015).

\bibitem{Kozik2015}
E. Kozik, M. Ferrero, and A. Georges, Nonexistence of the Luttinger-Ward Functional and Misleading Convergence of Skeleton Diagrammatic Series for Hubbard-Like Models, Phys. Rev. Let. {\bf 114}, 156402 (2015).

\bibitem{Balzer2011}
M. Balzer, N. Gdaniec, and M. Potthoff, Krylov-space approach to the equilibrium and nonequilibrium single-particle Green's function, J. Phys.:  Condens. Matter, {\bf 24},  035603  (2011).

\bibitem{Nocera2018}
A. Nocera, F.~H.~L. Essler, and A.~E. Feiguin, Finite-temperature dynamics of the Mott insulating Hubbard chain, Phys. Rev. B {\bf 97},  045146 (2018).

\bibitem{Chiara2018}
G. De Chiara and A. Sanpera, Genuine quantum correlations in quantum many-body systems: a review of recent progress, Rep. Prog. Phys. {\bf 81}, 074002 (2018).

\bibitem{Luiser2009}
M.~Luiser and G.~Kilmeck, Atomistic full-band simulations of silcon nanowire transistors: Effects of electron-phonon coupling, Phys. Rev. B {\bf 80}, 155430 (2009).

\bibitem{Rhyner2014}
R.~Rhyner and M.~Luiser, Atomistic modeling of coupled electron-phonon transport in nanowire transistors, Phys. Rev. B {\bf 89}, 235311 (2014).

\bibitem{Cavassilas2016}
N. Cavassilas, F.~Michelini, and M.~Bescond, On the local approximation of the electron-photon self-energy, J. Comput. Electron. {\bf 15}, 1233 (2016).

\bibitem{Bescond2018}
M.~Moussavou, M.~Lannoo, N. Cavassilas, D.~Logoteta, and M.~Bescond, Physically based diagonal treatment of the self-energy of polar optical phonons: Performance assessment  of III-V double-gate transistors, Phys. Rev. Appl. {\bf 10}, 064023 (2018).

\bibitem{Woolsey1970}
R.~B.~Woolsey and R.~M.~White, Electron-magnon interaction in Ferromagnetic semiconductors, Phys. Rev. B, {\bf 1}, 4474 (1970).

\bibitem{Frederiksen2004}
T. Frederiksen, {\em Inelastic electron transport in nanosystems}, M.S. thesis, Technical University of Denmark (2004).

\bibitem{Wells2019}
D. Wells and H. Quiney, A fast and adaptable method for high accuracy integration of the time-dependent Schr\"{o}dinger equation, Sci. Rep. {\bf 9}, 782 (2019).

\bibitem{Quirion2020}
D. Lachance-Quirion, S.~P.~Wolski, Y.~Tabuchi, S.~Kono, K.~Usami, Y.~Nakamura, Entanglement-based single-shot detection of a single magnon with a superconducting qubit, Science {\bf 367}, 425 (2020).

\bibitem{Fukuhara2013}
T. Fukuhara, P. Schau{\ss}, M. Endres, S. Hild, M. Cheneau, I. Bloch, and  C. Gross,  Microscopic observation of magnon bound states and their dynamics, Nature {\bf 502}, 76 (2013).

\bibitem{Morimae2005}
T.~Morimae, A. Sugita, and A. Shimizu, Macroscopic entanglement of many-magnon states, Phys. Rev. A {\bf 71}, 032317 (2005).

\bibitem{Haque2010}
M. Haque, Self-similar spectral structures and edge-locking hierarchy in open-boundary spin chains, Phys. Rev. A {\bf 82}, 012108 (2010).

\bibitem{Marini2018}
A. Marini and Y. Pavlyukh, Functional approach to the electronic and bosonic dynamics of many-body systems perturbed with an arbitrary strong electron-boson interaction, Phys. Rev. B {\bf 98}, 075105 (2018).

\bibitem{Siegrist2019}
F. Siegrist {\em et al.}, Light-wave dynamic control of magnetism, Nature {\bf 570}, 240 (2019).

\bibitem{White2004}
S. R. White and A. E. Feiguin, Real-time evolution using the density matrix renormalization group, Phys. Rev. Lett. {\bf 93}, 076401 (2004).

\bibitem{Schmitteckert2004}
P. Schmitteckert, Nonequilibrium electron transport using the density matrix renormalization group method, Phys. Rev. B {\bf 70}, 121302(R) (2004).

\bibitem{Daley2004}
A. J. Daley, C. Kollath, U. Schollw{\"o}ck, and G. Vidal, Time-dependent density-matrix renormalization-group using adaptive effective Hilbert spaces,  J. Stat. Mech: Theor. Exp. P04005 (2004).

\bibitem{Feiguin2011}
A. E. Feiguin, The density matrix renormalization group and its time-dependent variants, AIP Conf. Proc. {\bf 1419}, 5 (2011).

\bibitem{Stoudenmire2012}
E. M. Stoudenmire and S. R. White, Studying two-dimensional systems with the density matrix renormalization group, Annu. Rev. Condens. Matter Phys. {\bf 3}, 111 (2012).

\bibitem{Bertrand2019}
C. Bertrand,  O. Parcollet, A. Maillard, and X. Waintal, Quantum Monte Carlo algorithm for out-of-equilibrium Green's functions at long times, Phys. Rev. B {\bf 100}, 125129 (2019).

\bibitem{Bertrand2019a}
C. Bertrand, S. Florens, O. Parcollet, and X. Waintal, Reconstructing nonequilibrium regimes of quantum many-body systems from the analytical structure of perturbative expansions, Phys. Rev. X {\bf 9}, 041008 (2019).

\bibitem{Gaury2014}
B. Gaury, J. Weston, M. Santin, M. Houzet, C. Groth, and X. Waintal, Numerical simulations of time-resolved quantum electronics, Phys. Rep. {\bf 534},  1  (2014).

\bibitem{Bajpai2019b}
U.~Bajpai, B.~S.~Popescu, P. P.~Plech\'{a}\v{c}, B. K. Nikoli\'{c}, L.~E.~F.~F.~Torres, H.~Ishizuka, and N.~Nagaosa, Spatio-temporal dynamics of shift-current quantum pumping by femtosecond light pulse, J. Phys. Mater. \textbf{2}, 025004 (2019).

\bibitem{Schlosshauer2005}
M.~Schlosshauer, Decoherence, measurement problem, and the interpretations of quantum mechanics, Rev. Mod. Phys. \textbf{76}, 1267 (2005).

\bibitem{Mera2016}
H. Mera, T. G. Pedersen, and B. K. Nikoli\'{c}, Hypergeometric resummation of self-consistent sunset diagrams for electron-boson quantum many-body systems out of equilibrium, Phys. Rev. B {\bf 94}, 165429 (2016).


\bibitem{Frederiksen2007}
T. Frederiksen, M. Paulsson, M. Brandbyge, and A.-P. Jauho, Inelastic transport theory from first principles: Methodology and application to nanoscale devices, Phys. Rev. B {\bf 75}, 205413 (2007).

\bibitem{Lee2009}
W.~Lee, N.~Jean, and S.~Sanvito, Exploring the limits of the self-consistent Born approximation for inelastic electronic transport, Phys. Rev. B \textbf{79}, 085120 (2009).

\bibitem{phys814}
\emph{PHYS814: Advanced Quantum Mechanics}, \url{https://wiki.physics.udel.edu/phys814/}

\end{thebibliography}

\end{document}